\documentclass[12pt]{article}
\usepackage{amsmath}
\usepackage{graphicx,psfrag,epsf}
\usepackage{enumerate}
\usepackage{enumitem}
\usepackage{natbib}
\usepackage{url} 
\usepackage{xcolor}
\usepackage{caption}
\usepackage{subcaption}   
\usepackage{authblk}

\newcommand{\blind}{0}
\begin{document}

\def\spacingset#1{\renewcommand{\baselinestretch}%
{#1}\small\normalsize} \spacingset{1}


\if0\blind
{
  \title{\bf Estimands and their Estimators for Clinical Trials 
  Impacted by the COVID-19 Pandemic:\\ A Report from the NISS Ingram Olkin Forum Series on Unplanned Clinical Trial Disruptions}
\author[1, 2]{Kelly Van Lancker}
\author[3]{Sergey Tarima}
\author[4]{Jonathan Bartlett}
\author[5]{Madeline Bauer}
\author[6]{Bharani Bharani-Dharan}
\author[7,8]{Frank Bretz}
\author[9]{Nancy Flournoy}
\author[2]{Hege Michiels}
\author[4]{Camila Olarte Parra}
\author[10]{James L Rosenberger}
\author[11]{Suzie Cro}

\affil[1]{Department of Biostatistics, Johns Hopkins Bloomberg School of Public Health, Baltimore, U.S.A.}
\affil[2]{Department of Applied Mathematics, Computer Science and Statistics, Ghent University, Ghent, Belgium}
\affil[3]{Division of Biostatistics, Medical College of Wisconsin, U.S.A.}
\affil[4]{Department of Mathematical Sciences, University of Bath, U.K.}
\affil[5]{Division of Infectious Diseases, Keck School of Medicine, University of Southern California (ret), Los Angeles, U.S.A.}
\affil[6]{Novartis Pharmaceuticals, East Hanover, U.S.A.}
\affil[7]{Novartis Pharma AG, Basel, Switzerland}
\affil[8]{Section for Medical Statistics, Center for Medical Statistics, Informatics, and Intelligent Systems, Medical University of Vienna, Vienna, Austria}
\affil[9]{Department of Statistics, University of Missouri (emerita), Columbia, U.S.A.}
\affil[10]{National Institute of Statistical Sciences, and Department of Statistics, Penn State University, University Park, PA 16802-2111 U.S.A.}
\affil[11]{Imperial Clinical Trials Unit, Imperial College London, U.K. }
  \maketitle
} \fi

\if1\blind
{
  \bigskip
  \bigskip
  \bigskip
  \begin{center}
    {\LARGE\bf Title}
\end{center}
  \medskip
} \fi

\bigskip
\begin{abstract}
The COVID-19 pandemic continues to affect the conduct of clinical trials 
globally. Complications may arise from pandemic-related operational challenges such as site closures, travel limitations and interruptions to the supply chain for the investigational product, or from health-related challenges such as COVID-19 infections. 
Some of these complications lead to unforeseen intercurrent events in the sense that they affect either the interpretation or the existence of the measurements associated with the clinical question of interest. In this article, we demonstrate how the ICH E9(R1) Addendum on estimands and sensitivity analyses provides a rigorous basis to discuss potential pandemic-related trial disruptions and to embed these disruptions in the context of study objectives and design elements. We introduce several hypothetical estimand strategies and review various causal inference and missing data methods, 
as well as a statistical method that combines unbiased and possibly biased estimators for estimation. To illustrate, we describe the features of a stylized trial, 
and how it may have been impacted by the pandemic. This stylized trial will then be re-visited by discussing the changes to the estimand and the estimator to account for pandemic disruptions.
Finally, we outline 
considerations for designing future trials in the context of 
unforeseen disruptions.
\end{abstract}

\noindent%
\vfill

\newpage
\spacingset{1.45} 
\section{Introduction}
\label{sec:intro}

The COVID-19 pandemic, also known as the coronavirus pandemic, is an ongoing global pandemic of coronavirus disease 2019 (COVID-19) which is caused by severe acute respiratory syndrome coronavirus 2 (SARS-CoV-2). The coronavirus family is known to cause illness in humans, from the common cold to more severe or even fatal diseases. The World Health Organization (WHO) declared a Public Health Emergency of International Concern regarding COVID-19 on 30 January 2020, and later declared a pandemic on 11 March 2020. 

The COVID-19 pandemic has a broad impact, affecting general society, economy, culture, ecology, politics, and other areas. In particular, it continues to affect the conduct of clinical trials of medical products globally. Complications may arise from, for example, quarantines, site closures, travel limitations, interruptions to the supply chain for the investigational product, or other considerations if site personnel or trial participants become infected with COVID-19. Some of these complications lead to unforeseen intercurrent events in the sense that they affect either the interpretation or the existence of the measurements associated with the clinical question of interest while others prevent relevant data being collected and result in a missing data problem, ultimately posing a risk of seriously compromising the integrity and interpretability of ongoing clinical trials; see \cite{meyer2020statistical} and \cite{akacha2020challenges} for detailed discussions.  
 
The ICH E9(R1) Addendum on estimands and sensitivity analyses \citep{ich2019} introduces a framework for ensuring that all aspects of trial design, conduct, analysis, and interpretation align with the trial objectives. It also provides a rigorous basis to discuss potential pandemic-related trial disruptions and to embed them in the context of trial
objectives and design elements. More specifically, the ICH E9(R1) Addendum introduces five strategies to address intercurrent events, of which three will be used throughout this paper: 
\begin{itemize}
    \item \textbf{Treatment Policy Strategy.} The occurrence of an intercurrent event is considered irrelevant in defining the treatment effect of interest: the value for the variable of interest is used regardless of whether the intercurrent event occurs.
    \item \textbf{Hypothetical Strategies.} A setting is envisaged in which the intercurrent event would not occur: the value of the variable to reflect the clinical question of interest is the (predicted) value the variable would have taken in the hypothetical scenario defined.
    \item \textbf{Composite Variable Strategy.} An intercurrent event is considered to be informative about the patient’s outcome and is incorporated into the definition of the variable, usually as an unfavourable outcome. The newly defined composite variable is then a trade-off between efficacy and the occurrence of the intercurrent event.
\end{itemize}
Following the ICH E9(R1) Addendum, the description of an estimand involves precise specifications of five key attributes (treatment, population, variable, intercurrent events, and population level summary) which
should be aligned with the clinical trial objectives. This includes the definition of all attributes and the suitable strategies for addressing  intercurrent events.

For most settings, the original trial estimand likely remains the same following the emergence of the COVID-19 pandemic. However, the unforeseen intercurrent events due to COVID-19 introduce ambiguity to the original trial questions and sponsors need to discuss how to account for them. No adaptation of the original estimand implicitly suggests a treatment policy approach for all unforeseen intercurrent events. However, in the presence of the unforeseen intercurrent events related to operational challenges due to the COVID-19 pandemic, it seems plausible to frame clinical questions by using a hypothetical estimand strategy \citep{meyer2020statistical, akacha2020challenges}. The relevance and acceptability of a hypothetical estimand strategy seems to be less clear for intercurrent events related to health status, such as death due to COVID-19 in a cardiovascular trial where death is an outcome of interest. It is critical that the targeted outcome can be reliably estimated with justifiable and plausible assumptions, including the conduct of appropriate sensitivity analyses. If this is not possible (e.g., due to limited understanding
of the disease or drug for the imputation of missing or hypothetical data), an alternative estimand should be chosen. 

Different hypothetical strategies could be considered and care must be taken that the envisaged scenario, in which the intercurrent event would not occur, is precisely described \citep{akacha2020challenges}. Here, we consider three possible hypothetical strategies, all in the absence of the administrative and operational challenges  caused  by  the  pandemic, where  interest  lies  in  the  treatment effect: (1) in a world where the COVID-19 outbreak starts during the trial and individuals can suffer from COVID-19 infections, (2) if all individuals were recruited in a world where COVID-19 already exists, and individuals can suffer from COVID-19 infections, and (3) in the absence of COVID-19, i.e., in a world where the disease does not exist.
The second estimand seems to be important for the conduct of future trials, as it is reasonable to assume that future medical practice may be slightly different compared to pre-pandemic times, even if the majority is vaccinated against the virus and immunity levels remain high.

Similar views have also been published by other authors.
 For example, when assessing the impact of COVID-19 on oncology clinical trials, \cite{degtyarev2020assessing} argued that the estimand framework introduced in the ICH E9(R1) Addendum provides a common language to discuss the impact of COVID-19 in a structured and transparent manner. \cite{cro2020four} presented a four-step strategy for handling missing outcome data in the analysis of randomized trials
 that are ongoing during a pandemic, with missing data arising due to  complications as listed above. They considered treatment effects for a `pandemic-free world' and a `world including a pandemic.' 

In this article we go beyond the existing literature and focus not only on estimands, but also on estimators, to handle the impact of the pandemic on clinical trials. The remainder of  this article is organized as follows. In Section~\ref{sec:exam}, we describe a stylized clinical trial in neuroscience, together with intercurrent events caused by the pandemic and how they may be handled. We outline different hypothetical estimands in 
Section~\ref{sec:estimands}. In Section \ref{sec:estimation}, we review various estimators for these estimands, and for each approach, we discuss which data need to be collected, and the underlying assumptions. We describe different considerations for designing future trials in the context of any unforeseen disruptions (e.g., emergence of a new pandemic) in Section~\ref{sec:future} and conclude with a discussion in Section~\ref{sec:recom}. 

This manuscript is the product of a working group formed from `Day Three:  Estimands and Missing Data' of the National Institute of Statistical Sciences (NISS) Ingram Olkin Forum Series on Unplanned Clinical Trial Disruptions which took place on September 1, 2020.  For more information on this scholarly activity visit the event website: \url{https://www.niss.org/news/ingram-olkin-forum-series-investigates-role-estimands-and-missing-data}.

\section{Motivating Example: A stylized trial in neuroscience with a longitudinal outcome}
\label{sec:exam}
 
In this section we describe the features of a stylized trial in neuroscience, and how it may have been impacted by the pandemic. This stylized trial will be discussed in more detail in subsequent sections by describing how changes to the estimand and the estimator account for pandemic disruptions.



Consider a double-blind, randomized controlled trial of a potentially disease-modifying trial drug (experimental treatment) versus placebo for a neuroscience indication. The scientific objective guiding the primary estimand compares the outcome measured on a continuous disease rating scale after 24 months of follow-up between the two treatments, e.g., Clinical Dementia Rating scale Sum of Boxes (CDR-SB) score in Alzheimer's disease.

The trial was initiated prior to the pandemic and the targeted treatment effect was an effect irrespective of any trial treatment interruptions, discontinuations for trial-drug related reasons or initiations of rescue treatment but assuming that no discontinuations for administrative reasons had occurred. Therefore, the primary estimand was  formally defined prior to the pandemic as follows:
\begin{itemize}
    \item The target population includes all patients as defined by the trial's inclusion/exclusion criteria.
    \item The primary variable is defined as 
    a continuous disease rating measured after 24 months of follow-up. 
    The disease rating scale is assessed every 3 months until the final visit at 24 months.   
    \item The treatment of interest is the randomized treatment (experimental treatment or placebo) administered every 3 months throughout the follow-up period of 24 months or until discontinuation due to reasons potentially associated with trial medication.
    \item The intercurrent events and their associated {\em strategies} are:
    \begin{enumerate}
        \item Withdrawal from randomized trial medication due to reasons potentially associated with trial medication: {\em treatment policy strategy.}
        \item Withdrawal from randomized trial medication due to reasons unrelated to trial medication (e.g., moving away from the trial site): {\em  hypothetical strategy.}   The hypothetical scenario of interest is if patients who withdrew from randomized trial medication due to reasons unrelated to trial medication had instead remained on their randomized treatment.
        \item Interruption of the randomized trial medication for any reason: {\em  treatment policy strategy.}
        \item Initiation of rescue treatment: {\em treatment policy strategy.}
        \item Death due to any cause: {\em composite strategy,} where we assign the worst possible value when someone dies.
    \end{enumerate}
    \item The summary measure is the difference in the means of the primary variable between the trial arms. 
\end{itemize}
\textbf{Impact of the pandemic:}
As the trial was impacted by the pandemic, the protocol had to be amended and the following intercurrent events were added due to the pandemic:
\begin{enumerate}
\item[6.] Withdrawal from or interruption of randomized trial medication due to pandemic-related administrative or operational reasons.
    \item[7.] Infections with the COVID-19 virus, receiving COVID-19 vaccinations or receiving treatments for COVID-19 symptoms.
\end{enumerate}

In trials with a relatively low number of COVID-19-related intercurrent events, trial stakeholders may be interested in interpreting the treatment effect estimated by following the original analysis plan that ignores the unforeseen intercurrent events due to COVID-19 (i.e., the same statistical analysis as originally planned aside from possible missing data issues caused by COVID-19 intercurrent events); see, for example, \cite{meyer2020statistical}.
In such trials, an estimand of interest might therefore correspond to the treatment effect in the trial that actually took place before and after the onset of the pandemic. Following the intention-to-treat (ITT) principle, this can be obtained by handling all COVID-19-related intercurrent events with the treatment policy strategy. In what follows, we will denote this treatment effect as `ITT Estimand'. 
Note that the treatment effect estimated by this `simple'  analysis does not necessarily estimate the population measure determined by the original estimand due to the new intercurrent events.

Handling administrative and operational related intercurrent events with the treatment policy strategy might not reflect the objective of the trial since we expect these extreme events to disappear in the future (i.e., once the pandemic is over). Specifically, as discussed in \cite{meyer2020statistical}, this will often not be of scientific interest because the conclusions would not generalize to a post-pandemic world (i.e., when the pandemic has been re-classified as endemic by the WHO).
It therefore seems more reasonable to assume that the objective of a trial relates to a world without major disruptions of the healthcare system (i.e., no administrative and operational challenges due to the pandemic) although the coronavirus itself still exists.
Such a trial objective could be addressed using a suitable hypothetical strategy. Taking all of the above into consideration, we describe in the following section different possible hypothetical estimands to account for the different COVID-19-related intercurrent events for this motivating example.

\section{Possible hypothetical estimands
}
\label{sec:estimands}
In this section we discuss various estimand strategies for pandemic-related intercurrent events. Specifically, we introduce and discuss various hypothetical strategies, which differ from each other in the postulated hypothetical scenarios and populations (see Section \ref{subsec:est_descr}). All estimands are formally introduced using the potential outcome notation at the end of this section (see Section \ref{subsec:est_not}).

\subsection{A high-level description of the estimands}\label{subsec:est_descr}
\textbf{Estimand 1.} One conceivable hypothetical scenario is where interest lies in a treatment effect for the overall patient population (i.e., population recruited before, during and after the pandemic) where from the time point of the COVID-19 outbreak individuals can suffer from COVID-19 infections, but in the absence of administrative and operational challenges caused by the pandemic. 
 In this hypothetical world, neither COVID-19-related lockdowns and administrative restrictions nor COVID-19-related withdrawals from the trial medication and nonadherence due to pandemic-related reasons are happening. Therefore, intercurrent events related to operational and administrative challenges due to the pandemic can be addressed with the hypothetical strategy described here. However, some individuals may get COVID-19 infection and others can get vaccinated or re-vaccinated. Such COVID-19 health-related intercurrent events, which can lead to withdrawal from or interruption of the trial medication, can be handled with the treatment policy strategy. 
 
 One of the key attributes of an estimand (as defined in the ICH E9(R1) Addendum) is its population, which is even more important during a dynamically changing epidemiological situation. The pandemic is an example of an event potentially triggering such population changes: depending on the disease, the population of patients in a trial prior to the outbreak of COVID-19 may be different from the population during the pandemic and may be expected to be different again from the population in future post-pandemic times. For example, it might be that elderly patients are less likely to attend the doctor's office during the pandemic, which shifts the demographics of recruited patients. 
This might complicate the interpretation and representativeness of the target population. In addition, the definition of Estimand 1 (and its estimate) will likely be affected by how much information is available at the time of COVID-19 outbreak, i.e., the time at which the trial started relative to the outbreak of COVID-19 and the trial's pattern of recruitment over calendar time.
Consider, for example, a trial where 90\% of the patients have their primary endpoint already recorded at the time of the COVID-19 outbreak. These 90\% of the patients' data were not affected by COVID-19; in particular, they could not have COVID-19 or a COVID-19-related intercurrent event. In this case, we anticipate that the estimate associated with Estimand 1 will likely differ only by a small amount from the treatment effect in a world where COVID-19 does not exist (i.e., Estimand 3 to be discussed below). Conversely, if a substantial number of patients are recruited after the pandemic or the entire trial started after the pandemic emerged, we then expect the estimate associated with Estimand 1 to be closer to the treatment effect if all individuals were recruited in a post-pandemic world (i.e., Estimand 2 to be discussed next). 

Arguably Estimand 1 is not entirely in the spirit of the ICH E9(R1) Addendum, since the estimand, in particular its population attribute, is implicitly (re)defined in retrospect, given when the trial in question took place in relation to the COVID-19 outbreak. 
Nevertheless, it could well be of interest for stakeholders to ask what would have been the outcome of interest in the overall patient population, in the absence of administrative challenges caused by the pandemic. 
Hence, it seems desirable to clearly articulate the treatment effect it is estimating. Thus, Estimand 1 is a tool for interpreting the results of data analyses under a disruption, but it cannot be used to prospectively design studies.
Instead, the remaining estimands below are defined to reflect  changes in the population across various time periods (along with other differences in the estimand attributes and strategies to handle the intercurrent events) and might therefore be more relevant from a scientific point of view as well as for practical reasons (e.g., when designing future studies). In particular, we will make a distinction between the population recruited before, during and after the pandemic. The `before pandemic' patient population corresponds to the population defined through the trial's inclusion/exclusion criteria prior to the onset of the pandemic (i.e., prior to March 2020, which was declared as the start of the pandemic by WHO). The `during pandemic' patient population is defined by the patients recruited during the pandemic, while the `after pandemic' patient population is defined by the patients recruited after the pandemic (i.e., where overall society and the healthcare system have adapted to COVID-19). To distinguish the latter two cohorts, we need a clear definition of what a `post-pandemic' world entails. For instance, once WHO has declared a date for the end of the pandemic, this population could be defined by the population of patients recruited after this date. However, at the moment, the definition of these periods is not straightforward. Moreover, the `during pandemic' period might by itself not have homogeneous impact on the trial because of the wavelike development of the pandemic in many areas.  We discuss these definitions in more detail in Section \ref{sec:recom}.

\textbf{Estimand 2.} It is reasonable to assume that some aspects of medical practice in the post-COVID-19 period may differ compared to the  pre-COVID-19 period, as a consequence of the current pandemic and technological advances, even if the majority is vaccinated against the virus. For example, the adoption of telehealth, i.e., the use of digital and communication technologies to access health care services remotely, might increase substantially and this could make trial enrollment more attractive for subjects who are based in more rural areas or for more frail subjects who are reluctant to travel frequently to study sites. 
This could make the evidence collected before and during the pandemic less relevant for advising future clinical practice. Thus, a conceivable hypothetical scenario is where interest lies in the treatment effect in a patient population in a post-pandemic world 
(i.e., where overall society and the healthcare system have adapted to COVID-19). Note that this is a rather `practical' definition and variations of this population are possible (e.g., reflecting varying levels of vaccination/infections/... at the time of recruitment). As mentioned before, a reasonable possibility in the future might be to define this population based on the date for the end of the pandemic that will be set by WHO. 
As before, we consider the hypothetical setting where patients can suffer from COVID-19 infections but in the absence of administrative and operational challenges. This estimand differs from Estimand 1 due to the difference in the population attribute,
as we use the overall patient population (i.e., population before, during and after the pandemic) in Estimand 1 and the post-pandemic patient population in Estimand 2. Consequently, in Estimand 2 all patients could be affected by COVID-19 (e.g., they can get infected by COVID-19).

For Estimand 2, COVID-19 health-related intercurrent events can be addressed with the treatment policy strategy, whereas the intercurrent events related to operational and administrative challenges due to the pandemic can be addressed with the hypothetical strategy described above. 

\textbf{Estimand 3.} An alternative conceivable hypothetical scenario is the estimand associated with the original trial objective. This is the scenario where interest lies in the treatment effect in the absence of COVID-19, that is, in a world where the disease does not exist. Consequently, all COVID-19-related events should then be handled with the hypothetical strategy. In particular, the hypothetical scenario of interest is if patients who withdrew from randomized trial medication due to pandemic-related reasons had instead remained on randomized treatment. Also if patients who had an interruption to trial medication due to pandemic-related reasons had instead remained on randomized treatment without interruption. Finally, patients who were infected with COVID-19 virus, treated for or vaccinated against COVID-19 were not infected, treated or vaccinated.
In defining this estimand, we focus on the patient population actually enrolled during the pre-pandemic period. Specifically, estimation of this estimand (see Section \ref{sec:estimation}), 
will require that the sample population facilitates estimation of the treatment effect (as defined by Estimand 3) in the patient population actually enrolled during the pre-pandemic period.
Note that in practice sponsors may be facing a shift in the population due to the pandemic, leading to challenges in estimating this estimand; see Section \ref{sec:recom} for a detailed discussion.

Throughout the rest of this article we will focus on estimating the above (hypothetical) estimands. Note, however, that other estimands may be of interest that take into consideration the unforeseen COVID-19-related intercurrent events; see Section \ref{sec:recom}. 

\subsection{Notation and formalizing the hypothetical estimands}\label{subsec:est_not}
Following the ICH E9(R1) Addendum, we define the estimands with respect to their five attributes: population, treatment, variable (endpoint), intercurrent events and a population summary measure. The population attribute of the estimand includes not only a patient population but also a state of the pandemic. In the neuroscience example (see Section \ref{sec:exam}), the \textbf{population} ($\Omega$) is the patient population satisfying the inclusion/exclusion criteria for the targeted disease population (e.g., in Alzheimer's disease) as defined in the trial protocol. At the design stage, the plan is to recruit a (random) sample of $n$ patients from $\Omega$. Index $i$ ($1, \dots, n$) is used to enumerate the enrolled patients. The patients are randomized to the intervention and control groups; the \textbf{treatment} indicator is $A$, where $A=1$ if the patient received the experimental treatment and $0$ if they received the placebo control. 
The baseline covariate data, $\mathbf{X}_{0}$, typically include patient characteristics, lab tests, and other assessments conducted before randomization.
To be able to relax the assumptions when estimating the different estimands (see Section \ref{sec:estimation}), we allow the covariate data to change over time by using a subscript $t$: $\mathbf{X}_{t}$, $t\geq0$. The \textbf{endpoint} $Y_{t}$ is the primary outcome measure (e.g., CDR-SB for Alzheimer’s disease), which can be measured at various time points $t$. In the neuroscience example, the assessment is completed every three months, so that $Y_{t}$ is measured up to nine times $(t=0,\ldots,8)$. The primary endpoint then corresponds to $Y_8$. The \textbf{intercurrent events} will be quantified by indicators $\mathbf{C}_{t}$, where $\mathbf{C}_{t} = 1$ indicates an intercurrent event that occurred (between time $t-1$ and $t$) that potentially affects endpoint variables and possibly time-varying covariates from visit $t$ on, $\mathbf{C}_{t} = 0$ otherwise. In the above characterization, $\left(A, \mathbf{X}_t, \mathbf{C}_t, Y_t\right)$ are random  variables but observable over time. Indexing of the individual elements of multivariate random variables $\mathbf{X}_t$ and $\mathbf{C}_t$ will be done using a superscript in parentheses. For the neuroscience example, the five pandemic-unrelated (1-5) and two pandemic-related intercurrent events (6-7) will be the elements of $\mathbf{C}_t=\left(C_{t}^{(1)}, \ldots, C_{t}^{(7)}\right)$, as numbered in Section \ref{sec:exam}. 
The estimation target is a \textbf{population summary measure} $\theta$. The sample will be distinguished from population random variables by the presence of index $i$. Then, $\left(A_{i}, \mathbf{X}_{it}, \mathbf{C}_{it}, Y_{it}\right)$ represents sample data for subject $i$   at visit $t$ ($t=0,\ldots,8$). 

For completeness, we first define the estimand corresponding to the intention-to-treat effect discussed at the end of Section \ref{sec:exam} (i.e., ITT estimand):
\begin{align}\label{itt}
\begin{split}
\theta_{ITT} &= E_{\Omega_{overall}}\left\{Y_{8}(A=1)\right\} - E_{\Omega_{overall}}\left\{Y_{8}(A=0)\right\},
\end{split}
\end{align} 
where $Y_8(A=k)$ is the counterfactual/potential outcome at time point $t=8$ under treatment $A=k$ ($k\in\{0,1\}$), that is, the outcome that an individual would have experienced (at time $t=8$) if he/she had received treatment $A=k$ \citep{lipkovich2020causal}. Moreover, $\Omega_{overall}$ denotes the overall patient population before, during and after the pandemic, as a whole. That is, $\Omega_{overall}=\Omega_{prior}\cup\Omega_{during}\cup\Omega_{after}$, where $\Omega_{prior}$, $\Omega_{during}$ and $\Omega_{after}$ denote the patient population prior to the onset of the pandemic, during the pandemic and after the pandemic, respectively.

\textbf{Estimand 1} is the effect of the treatment for the overall patient population where, from the time point of the COVID-19 outbreak, individuals can suffer from COVID-19 infections, but in the absence of administrative and operational challenges caused by the pandemic:
\begin{align}\label{estimand1}
\begin{split}
\theta_1 &= E_{\Omega_{overall}}\left\{Y_{8}(A=1,C_t^{(6)}=0, \forall t)\right\} - E_{\Omega_{overall}}\left\{Y_{8}(A=0,C_t^{(6)}=0, \forall t)\right\},
\end{split}
\end{align} 
where $Y_8(A=k,C_t^{(6)}=0, \forall t)$ is the counterfactual outcome at time point $t=8$ under treatment $A=k$ ($k\in\{0,1\}$) and where we intervene (somehow) to set the relevant COVID-19-related intercurrent events to zero at all time points. How this can be implemented by predicting patient trajectories is described in Section \ref{sec:estimation}. In this hypothetical world, neither COVID-19-related lockdowns and administrative restrictions nor COVID-19-related withdrawals from randomized trial medication and nonadherence due to pandemic-related reasons are happening. 

\textbf{Estimand 2} is the effect of the treatment in a post-pandemic patient population $\Omega_{after}$, where individuals can suffer from COVID-19 infections but in the absence of administrative and operational challenges caused by the pandemic:
\begin{align}\label{estimand2}
\begin{split}
\theta_2 &= E_{\Omega_{after}}\left\{Y_{8}(A=1,C_t^{(6)}=0, \forall t\right\}  - E_{\Omega_{after}}\left\{Y_{8}(A=0,C_t^{(6)}=0, \forall t)\right\}.
\end{split}
\end{align} 

\textbf{Estimand 3} is the treatment effect in the absence of COVID-19, that is, in a world where the disease does not exist. It is defined as
\begin{align}\label{estimand3a}
\begin{split}
\theta_{3} &= E_{\Omega_{prior}}\left\{Y_{8}(A=1,C_t^{(j)}=0, \forall t,j=6, 7)\right\}  - E_{\Omega_{prior}}\left\{Y_{8}(A=0,C_t^{(j)}=0, \forall t,j=6, 7)\right\}.
\end{split}
\end{align} 
Here, $Y_8(A=k,C_t^{(j)}=0, \forall t,j=6, 7)$ is the counterfactual outcome at time point $t=8$ under treatment $A=k$ ($k\in\{0,1\}$) and where we intervene (somehow) to set the relevant COVID-19-related intercurrent events (treatment withdrawal/interruptions, infections) to zero at all time points.

In Table \ref{tab:differences} we summarize the main differences between the estimands introduced above with respect to their attributes.
\begin{table}[h!]
\centering
 \caption{\label{tab:differences} Main differences between various estimands with respect to the estimand attributes. Overall population consists of the population before COVID-19 outbreak, population during the pandemic and population after the pandemic.}
 \begin{tabular}{l c c c} 
 \hline
Estimand & Population & Strategy for $C_t^{(6)}$  & Strategy for $C_t^{(7)}$\\
 \hline
 ITT Estimand & Overall  & Treatment policy & Treatment policy\\
 Estimand 1 & Overall & Hypothetical & Treatment policy\\
 Estimand 2 & After pandemic & Hypothetical & Treatment policy\\
 Estimand 3 & Prior to COVID-19 outbreak & Hypothetical & Hypothetical  \\
 \hline
 \end{tabular}
\end{table}

\section{Potential estimators for hypothetical estimands in ongoing trials}\label{sec:estimation}
In this section, we discuss potential estimators for the different (hypothetical) estimands outlined in Section \ref{sec:estimands}. More specifically, we review in Section~\ref{sec:missing} various estimation methods using approaches for handling missing data, such as likelihood-based analysis, multiple imputation, as well as simple and augmented inverse probability weighting. In Section~\ref{sec:est_combining} we then review methods for combining unbiased with possibly biased estimators.  

\subsection{Estimation using methods for handling missing data}
\label{sec:missing}
\begin{figure}[ht]
	\centering
	\includegraphics[width=1\textwidth]{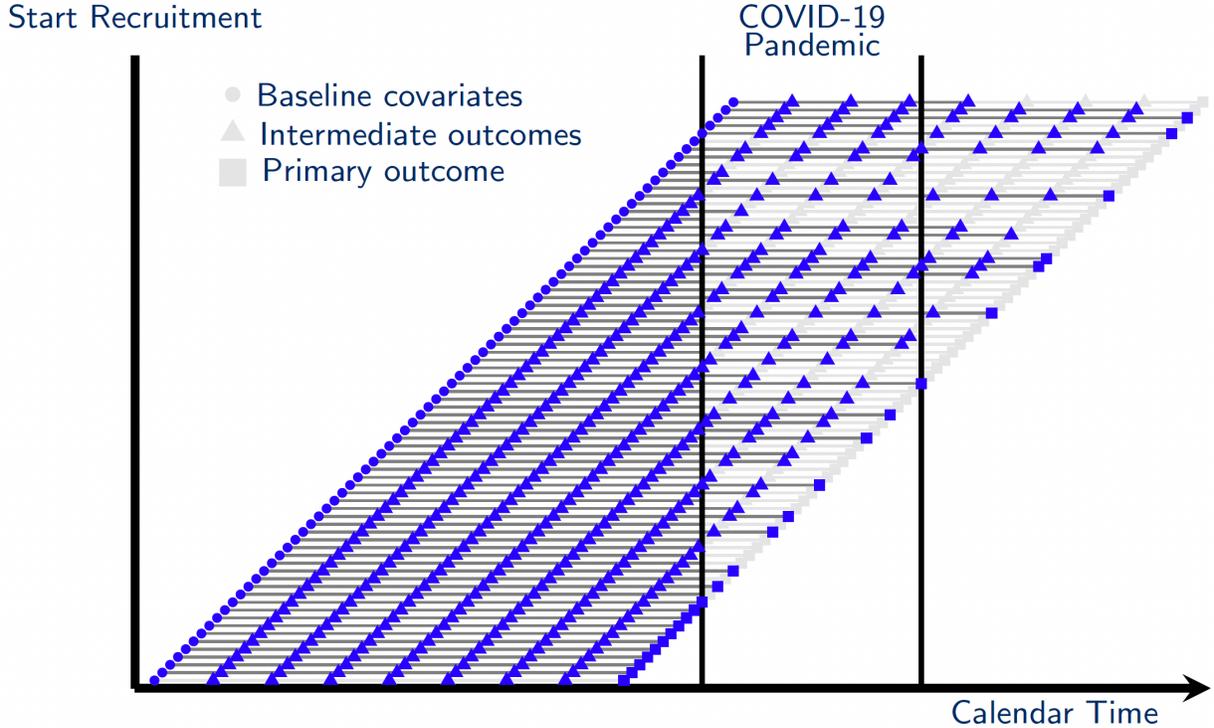}
	\caption{
		Visualization of the available data. The blue symbols show the available information (for the estimand of interest), while the gray symbols show the missing information (for the estimand of interest). Note that data observed after a relevant intercurrent event (for the estimand of interest), may be physically missing or, if observed, can be initially set missing.
	}
	\label{fig:covid_impacted}
\end{figure}

We first consider estimation 
using methods traditionally employed to handle missing data \citep[see e.g.,][]{cro2020four, parra2021hypothetical}. In such settings, patient data, after relevant (hypothetically handled) intercurrent events may either be physically missing or, if observed, can be treated as missing as these data are (usually) not relevant in estimating the estimand of interest. We note though that when estimating hypothetical estimands it is possible to exploit data observed after the intercurrent event via G-computation \citep{parra2021hypothetical}.  We assume a monotone (i.e., not intermittent) missingness pattern for patients experiencing relevant intercurrent events (for the estimand of interest) (see Figure \ref{fig:covid_impacted}).  This differs from the approach in Section \ref{sec:est_combining} where data observed after intercurrent events (i.e., data that is not physically missing) may be used in the estimation if the treatment effect is not affected. In practice, a trial may also have interim missing data or data missing for other reasons that are not due to intercurrent events, however, for simplicity, in the following discussion we first assume this is not the case.

For the neuroscience trial, for Estimand 1 (treatment effect in the whole patient population where patients can suffer COVID-19 from the point of the pandemic, but in the absence of administrative/operational pandemic challenges), only patient data observed post-intercurrent event $C_{t}^{(6)}$ would be set missing. For Estimand 2 (treatment effect in a patient population after the pandemic, where patients can experience COVID-19 infections but in absence of administrative/operational pandemic challenges); 
any data that was observed after experiencing intercurrent events of type $C_{t}^{(6)}$ can be set missing. Assuming no shift in population (i.e., the overall sample is representative of the targeted post-pandemic population), data from patients recruited before and during the pandemic can be retained and used in addition to data from the patients recruited after the pandemic, whilst pre- and during-pandemic patients should be removed entirely when we want to allow for the possibility of a shift in population. Note that using data from pre- and during-pandemic patients before experiencing intercurrent events of type $C_{t}^{(6)}$ implicitly assumes that the treatment effect is not changing over time (i.e., not changing because of the pandemic). In trials where this is not a reasonable assumption, all (post-baseline) follow-up data (including those collected in the post-pandemic period) for participants recruited pre- and during-pandemic should also be set to missing.
For Estimand 3 (treatment effect in the absence of COVID-19), any patient data observed after experiencing intercurrent events $C_{t}^{(6)}$ (treatment withdrawal/interruptions) or $C_{t}^{(7)}$ (COVID-19 health-related intercurrent events) should be set to missing.
In addition, assuming no shift in population, data from all patients can be included, whilst participants recruited after the onset of the pandemic would have to be removed entirely when we want to allow for the possibility of a shift in population.

Formally, for each participant $i$ ($i\in \{1, \dots, n\}$) let $D_i$ denote the last observation time in the absence of a relevant hypothetically handled intercurrent event, where $D_i$ can take values $t=0,...,8$. $D_i=8$ if no hypothetical intercurrent events occur. For each patient let the column vector $\mathbf{Y_{Oi}}=(Y_{i0},...,Y_{iD_{i}})^T$ denote the patient's observed responses at times $t=0,...,D_{i}$, and, if $D_i<8$, let the column vector $\mathbf{Y_{Pi}}=(Y_{i(D_{i}+1)},...,Y_{i8})^T$ denote actual (as opposed to counterfactual hypothetical) post-intercurrent responses at times $D_{i}+1,...,8$ (which may or may not be observed). The responses we are targeting in the hypothetical scenario is the vector $\mathbf{Y_{Mi}}$ of missing hypothetical responses at times $D_{i}+1,...,8$. Let  $\mathbf{Y_{M}}$ denote the column vector of missing values $(\mathbf{Y_{M1}^T},...,\mathbf{Y_{Mn}^T})^T$ which are the responses in the hypothetical scenario we are targeting and let $\mathbf{Y_{O}}$ denote the column vector of observed data relevant for the estimand $(\mathbf{Y_{O1}^T},...,\mathbf{Y_{On}^T})^T$.

Methods for estimation commonly used in missing data situations that make plausible assumptions for the post-intercurrent event data, $\mathbf{Y_{M}}$, that align with the hypothesised scenario/estimand of interest can then be used. Different hypothetical estimands will naturally require different assumptions for post-intercurrent event data.  Therefore, the most appropriate missing data method will depend on the hypothetical scenario underlying the estimand of interest, and the corresponding required assumptions for the unobserved post-intercurrent event data. Careful thought is required as to whether information can be borrowed from those observed in the trial to predict the hypothetical measurements of interest. If the relevant hypothetical post-intercurrent event data is postulated to be similar to the remaining observed data in the trial conditional on baseline covariates and/or earlier response data in some form, a \emph{missing at random} (MAR) assumption may be appropriate for estimation (see e.g., \cite{cro2020four, meyer2020statistical}). As originally described by \cite{rubin76}, data are MAR when the probability of missingness does not depend on the values of the unobserved outcomes, given observed covariates (possibly time-varying) and outcomes being analysed. Or equivalently, the unobserved data has the same distribution as that observed, conditional on observed data (covariates and/or outcomes) within the analysis.

In the neuroscience trial for Estimand 1,  we are interested in the hypothetical scenario where 
patients had not withdrawn from or interrupted treatment due to pandemic-related reasons ($C_{t}^{(6)}=0, \forall t$). For post-intercurrent event data, a MAR assumption making use of treatment group, a pandemic time point indicator (patient observed pre/during or potentially post pandemic), covariates (possibly time-varying) and earlier observed responses, may be most relevant to assume for estimation. Specifically, such a MAR assumption implies that patients experiencing pandemic-related treatment withdrawal/interruptions had instead behaved like similar patients observed in the same treatment group who had not withdrawn or interrupted treatment due to pandemic-related reasons ($C_{t}^{(6)}=0, \forall t$). The inclusion of a pandemic time point indicator is necessary to ensure the missing post-intercurrent event data is modelled based upon the data of those observed during the appropriate time point. Moreover, for MAR to hold here, we need to ensure that we condition on common causes of the hypothetical intercurrent event (i.e., administrative and operational challenges due to COVID-19) and the outcome of interest, which possibly include --among others-- country, center and age.  Here, as $C^{(7)}$ is handled via the treatment policy strategy and all observed data are used in the analysis regardless of this event, hypothetical post-intercurrent event data for $C^{(6)}$ will thus be modelled on the data of patients in the same treatment group which includes the presence of this event. 

For Estimand 2, a MAR assumption, making use of treatment group, covariates (possibly time-varying and possibly including a pandemic time point indicator) and earlier observed responses, may be most relevant to assume for estimation. 
Under the assumption that there is no shift in population and that the treatment effect is not changing over time, MAR will allow us to model the unobserved patient data appropriately on all the data (of all patients) that were not impacted by operational and administrative challenges. When allowing for a shift in population and/or a change in the treatment effect over time (due to the pandemic), MAR will allow us to model the unobserved patient data appropriately on the data of only those participants recruited after the pandemic, and not impacted by operational challenges.

For Estimand 3, interest lies in the hypothetical scenario where patients 
did not get COVID-19 or had not withdrawn from or interrupted treatment due to pandemic-related reasons ($C_{t}^{(j)}=0, \forall t, j=6,7$). For post-intercurrent event data, a MAR assumption, making use of treatment group, covariates (possibly time-varying) and earlier observed responses, may be most relevant to  assume for estimation. Specifically, such a MAR assumption implies that patients experiencing COVID-19 or pandemic-related treatment withdrawal/interruptions had instead behaved like similar patients observed in the same treatment group who did not get COVID-19 or had withdrawn or interrupted treatment due to pandemic-related reasons ($C_{t}^{(j)}=0, \forall t, j=6,7$).
This MAR assumption will allow us to model the post-intercurrent event data based on the remaining observed data of those who were observed when COVID-19 did not exist, or, when COVID-19 existed, but were not impacted by it. Such an approach naturally requires a sufficient sample size of observed data for covariate combinations to reliably model the missing data.

A MAR assumption may not always be plausible for missing hypothetical responses, where patient data are alternatively supposed to be quite different from the observed patient data in some form in the hypothetical scenario, even after conditioning on pre-intercurrent event data. 
A corresponding \emph{missing not at random} (MNAR) assumption would be required for the post-intercurrent event data. Data are MNAR if the probability of missingness depends on the unobserved values in some form even after controlling for other observed variables. The distribution of the unobserved data will be different to that of the observed data, both marginally and conditionally on observed variables. 

Critically, any MAR missing data assumption is not testable and has to be assumed to be correct for estimating the treatment effect in the hypothetical scenario, i.e., under MAR that it is plausible to borrow information from the observed to predict the hypothetical measure of interest. MNAR data analyses would require alternative (also untestable) assumptions. Therefore sensitivity analysis under alternative plausible assumptions for the post-intercurrent event data is recommended.

For simplicity in the following discussion on subsequent methods that can be used to obtain inference under MAR or MNAR (as appropriate) we assume there is either no other missing data in the trial (i.e., missing data only occurs due to intercurrent events to be handled with a hypothetical strategy) or that any other missing data can be appropriately handled under the same assumption as post-intercurrent event data that are handled via the hypothetical scenario. 



\subsubsection{Likelihood-based analysis}

Consistent and asymptotically efficient parameter estimates under a MAR assumption can be obtained via a suitable likelihood-based analysis, provided the assumption about the missing data mechanism is met, the parameters of the data model and missingness mechanism are distinct such that the parameter space of the joint data and missingness distribution is the product of the parameter spaces of the data and missingness mechanism  (also known as variation independent) and the complete data model is correctly specified (\citeauthor{lilrub87} \citeyear{lilrub87}).
Here, with missing data we mean the missing hypothetical data. In general, a parametric model is assumed for the complete data  (including all variables within the MAR assumption) and inference is obtained based on the observed data likelihood. Therefore, when the aim is to estimate a hypothetical treatment effect in the absence of particular intercurrent events where patients are instead proposed to behave like those not experiencing the event (for example pre-treatment withdrawal/interruptions due to pandemic-related reasons for Estimand 1), a likelihood-based model suitable for the type of data at hand can be fitted to only the relevant pre-intercurrent event observed data $\mathbf{Y_{O}}$ when MAR is appropriate. 
In the neuroscience trial, a linear mixed model for repeated measures, including all covariates included in the MAR assumption, i.e., treatment, pandemic time point indicator and relevant baseline covariates (possibly also crossed with time if appropriate), can then be fitted to all observed data unaffected by relevant intercurrent events $\mathbf{Y_{O}}$, within a likelihood-based framework. This will estimate the treatment effect in the absence of the specified intercurrent events, under the assumption that patient data would be similar to those in their treatment group who are observed  and do not experience these events and who are similar with respect to the observed data (MAR).

Maximum likelihood estimation can also be applied to time-to-event outcomes, e.g., using a Cox model fit within a (partial) likelihood-based framework. 
In other settings with a categorical, ordinal or count outcome, an alternative  appropriate generalised mixed model may be used.

Another analysis option under MAR, where the observed data likelihood is central is the Bayesian approach. Bayesian inference assigns a prior distribution to the parameters of the data model and the observed data likelihood is multiplied by the prior to form the posterior distribution for inference. For more details on Bayesian methods for incomplete data see Chapter 5 in \cite{Molenberghs/Fitzmaurice/Kenward:2014}.  


\subsubsection{Multiple Imputation (MI)}

Multiple Imputation (MI) is an alternative method for consistent and asymptotically efficient estimation when a MAR assumption is appropriate for the hypothetical post-intercurrent event data, provided the assumption about the missing data mechanism is met
and models used for imputation and analysis are correctly specified (\citeauthor{rubin87}, \citeyear{rubin87}). 
MI involves sampling the missing data, $\mathbf{Y_{Mi}}$, from the conditional distribution of the missing responses given the observed responses and baseline covariates (potentially including a pandemic time point indicator if relevant for the estimand at hand), $f(\mathbf{Y_{Mi}}|\mathbf{Y_{Oi}},\mathbf{X_{i0}},\mathbf{\eta})$, within a Bayesian framework where $\mathbf{\eta}$ are the parameters of this distribution. The missing data are sampled multiple times to create multiple completed datasets. Post-imputation, the analysis model of interest is fitted to each of the imputed data sets. Results across the imputed datasets are then combined using Rubin’s rules (\citeauthor{rubin87}, \citeyear{rubin87}) for inference. 
Rubin's rules will only give valid standard errors if the imputations are `proper' and allow for uncertainty in the parameters of the imputation model (\citeauthor{rubin87}, \citeyear{rubin87}). Additionally, the imputation model should include all the variables included in the analysis model and in the correct functional form (e.g., including any interactions that are required). Otherwise if the imputation model omits one or more variables included in the analysis model then Rubin’s rules will not provide valid inference.

When the imputation and analysis models align then multiple imputations (MI) is expected to give the same results as an observed likelihood-based analysis (within Monte Carlo sampling variability). However, an appealing feature of MI is that additional variables that are thought to be predictive of missingness and outcome but are not required in the analysis model, referred to as auxiliary variables, can also be included in the imputation model. This can help strengthen the validity of the underlying MAR assumption and can be more efficient with precision increased (\citeauthor{collins:2001}, \citeyear{collins:2001}, \citeauthor{zhou:2001}, \citeyear{zhou:2001}). 
For example, in the neuroscience trial for Estimand 3, suppose the presence of a co-morbidity (e.g., diabetes) was identified as a predictor of outcome and  also associated with interruption of trial medication due to COVID-19 as patients with that co-morbidity missed treatment visits as they were at greater risk of severe COVID-19 disease; but the treatment effect was not required to be estimated conditionally on that co-morbidity. MI can be applied to the relevant observed data, $\mathbf{Y_{O}}$, $\mathbf{X_0}$  and $A$, to impute all outcome data that is either missing or set missing post relevant intercurrent events, $\mathbf{Y_{M}}$, under the assumption of MAR conditional on treatment group, baseline covariates (including the co-morbidity) and earlier outcomes. Incorporating the co-morbidity  within the imputation model enables the unobserved data to be modelled conditionally on the co-morbidity. But co-morbidity does not need to be included in the analysis model. Care must however be taken with binary and categorical outcomes, since the inclusion of as many relevant variables as possible can increase the probability of perfect prediction and may consequently generate biased results. To approach perfect prediction further guidance is provided by \cite{white:2010}.

Up to here we have discussed the use of MI for estimation when MAR assumptions are appropriate for post-intercurrent event data. When post-intercurrent event hypothetical data is believed to be quite different to  observed data, 
an analysis that makes a MNAR assumption will be required. MI also provides an accessible analysis option for estimation under MNAR. For example, suppose individuals who have higher outcome values (e.g., higher CDR-SB scores) are more likely to get infected with COVID-19.  For Estimand 3, to handle this intercurrent event, a hypothetical scenario may be envisaged where the COVID-19 infection did not occur and the patients who became infected instead had a worse/higher outcome than those observed in the trial (who had lower outcome values).   Rather than imputing data based on the observed data distribution (i.e., under MAR), data can be imputed for the affected patients under an alternative MNAR distribution that aligns with the hypothetical scenario. When MI is performed in such a manner this has been termed `controlled MI' as the analyst has direct control over the imputation distribution (\citeauthor{Kenward:2015}, \citeyear{Kenward:2015}).  

Any MNAR distribution can be utilised within controlled MI. When a better or worse outcome is assumed for the hypothetical outcome, relative to that predicted based on the observed unaffected data (under MAR), an accessible principled approach involves altering the observed MAR imputation distribution by a specified numerical sensitivity parameter termed `delta', which in simple cases may be a scalar or in more complex cases a vector parameter. Formally, MNAR imputation entails sampling the missing data $\mathbf{Y_{Mi}}$ from  the conditional distribution, $f(\mathbf{Y_{Mi}}|\mathbf{Y_{Oi}},\mathbf{X_{i0}},\mathbf{\eta_D})$, within a Bayesian framework where $\mathbf{\eta_D}$ represents the parameters of this distribution whose values differ across missingness pattern $D$. For each imputation, $\mathbf{\eta_D}$ must be first formed to enable imputation using the standard MI procedure from the above conditional distribution. When a delta approach is adopted for MNAR imputation, $\mathbf{\eta_D}$ is constructed using the parameters of the MAR implied conditional distribution, $\mathbf{\eta}$, and a numerical sensitivity parameter. For each missing data pattern $D$, denote $\mathbf{\delta_{D}}$ as the postulated numerical difference in the parameter(s) governing the departure from MAR, which will create a shifted distribution for imputation. MI then proceeds using the shifted distribution. In the most simple case, e.g., with a single follow-up outcome, $\mathbf{\delta_{D}}=\mathbf{\delta}$ can be a single numerical parameter; for continuous outcomes this will govern the conditional mean difference between the MAR and MNAR distribution for the missing outcome; for binary outcomes this can govern the mean difference in the log odds. In more complex cases, e.g., with longitudinal follow-up, $\mathbf{\delta_{D}}$ can be a numerical vector parameter.   The numerical sensitivity parameter $\mathbf{\delta_{D}}$ may also vary by treatment arm.


In any trial setting, the extent to which the parameters of the distribution of missing data are likely to differ from the MAR parameters for each missing data pattern requires careful consideration and justification. The data cannot inform about the values of delta, therefore a sensitivity analysis exploring the impact of alternative delta values is recommended to explore the robustness of inferences. A commonly applied approach, a so‐called ‘tipping‐point’ analysis, involves increasing `delta' until the trial conclusions change. Such an approach may also be particularly useful for sensitivity analyses, following a primary MAR analysis. When the departure from MAR varies by treatment arm, i.e., the numerical sensitivity parameter $\mathbf{\delta_{D}}$  varies by treatment arm, there will generally be greater  sensitivity to different values of $\mathbf{\delta_{D}}$, i.e., larger differences between the resulting treatment effect and that observed under MAR, compared to when the departure from MAR is the same for all individuals \citep{Leacy:2017}.

Up to here we have assumed any other missing data in the trial can be handled under the same assumption as for the hypothetical post-intercurrent event data.  When different missing data assumptions are required for different groups of participants in the same trial analysis (e.g., MAR for general loss to follow-up and MNAR for hypothetical post-intercurrent event data), controlled multiple imputation provides an accessible tool for estimation. Imputation can proceed using different distributions for the missing data of different groups of participants. For a more detailed practical overview of controlled multiple imputation procedures, which includes worked examples where different assumptions are incorporated in one analysis, see \cite{Cro2020Practicalguide}.

The described likelihood-based analyses (e.g., mixed models) and MI methods make a relatively strong MAR assumption, which requires that no time-varying covariates are relevant, except the outcome, to render missingness independent of the missing hypothetical outcomes.
Although these methods can accommodate time-varying prognostic covariates of outcome that are associated with having an intercurrent event, it complicates the implementation as the (missing) covariates and outcomes then need to be jointly modeled/imputed. This can then increase the risk of model misspecification. In addition, when patients with and without missing data (i.e., patients with and without relevant intercurrent events) are very different, these methods rely on (strong) extrapolation. In the following sections, we therefore consider methods that do not rely on extrapolation and easily allow for time-varying prognostic factors of missingness \citep{Robins00marginalstructural, hernan2001marginal, hernan2020}.

\subsubsection{Inverse Probability Weighting (IPW)}

A different approach to handle missing data is to weight the observed outcome data in an appropriate manner that corrects for the patients with missing outcome data. This class of estimators directly uses models for the missingness mechanism itself \citep{Robins00marginalstructural, hernan2001marginal, hernan2020}:

\begin{enumerate}
			\item Under MAR, the probability of missingness (i.e., due to a relevant intercurrent event) can be estimated based on the observed (longitudinal) outcomes, time-varying covariates and any additional auxiliary variables that are thought to predict missingness.
			Depending on the estimand, these might include a pandemic time point indicator (patient observed pre/during or potentially post pandemic). Denote by $\mathbf{\bar{C}_{t-1}}$, $\mathbf{\bar{X}_{t-1}}$ and $\bar{Y}_{t-1}$ the history until (and including) time point $t-1$ of the intercurrent events, the covariates and outcomes, respectively. Moreover, let $C^*_t$ equal $0$ if $C^{(6)}_t=C^{(7)}_t=0$ and $1$ otherwise for Estimand 3 (see Equation \eqref{estimand3a}), and let $C^*_t$ equal $0$ if $C^{(6)}_t=0$ and $1$ otherwise for Estimand 1 (see Equation \eqref{estimand1}) and Estimand 2 (see Equation \eqref{estimand2}). Then, at each time point $t\in\{1, \dots, 8\}$, we estimate the conditional probability  
$$P(C^*_t=0|A, \mathbf{\bar{C}_{t-1}}, \mathbf{\bar{X}_{t-1}}, \bar{Y}_{t-1})$$
by fitting a logistic regression in the patients with $\mathbf{C}^*_{t-1}=\mathbf{0}$, where $\mathbf{0}$ is a vector with only zeros and $\mathbf{C}^*_{t-1}=(C^*_{1}, \dots, C^*_{t-1})$.
			\item These probabilities can then be used to calculate the inverse probability weights, $$W_i=\prod_{t=1}^{D_i}\frac{1}{P(C^*_{i,t}=0|A_{i}, \mathbf{\bar{C}_{i,t-1}}, \mathbf{\bar{X}_{i,t-1}}, \bar{Y}_{i,t-1})}$$
			for each patient $i$.
		\item These weights can then be used to weight the complete cases and obtain an estimate of the treatment effect as follows:
\begin{align*}
&n_1^{-1}\sum_{i=1}^nI(A_i=1, \mathbf{C}^*_{i,8}=\mathbf{0})W_iY_{i,8}-n_0^{-1}\sum_{i=1}^nI(A_i=0, \mathbf{C}^*_{i,8}=\mathbf{0})W_iY_{i,8}
\end{align*}
with $n_1$ and $n_0$ denoting the number of (relevant) patients in the experimental treatment and placebo arm, respectively. Here, with `relevant' we mean the patients that are considered relevant for the estimation of the considered estimand. Alternatively, we can use stabilized weights by replacing $n_1$ and $n_0$ by $\sum_{i=1}^nI(A_i=1, \mathbf{C}^*_{i,8}=\mathbf{0})W_i$ and $\sum_{i=1}^nI(A_i=0, \mathbf{C}^*_{i,8}=\mathbf{0})W_i$, respectively.
		\end{enumerate}

Each complete case's outcome data is weighted by the reciprocal of the probability of that patient having observed outcome data on the basis of his/her characteristics. Weighting thus corrects for the selection bias (in `sufficiently’ large samples), induced by the fact that patients with complete data may form a selective subgroup, when all prognostic factors of missingness have been measured (i.e., under MAR). Intuitively, this is because IPW reconstructs a `pseudo’ random sample from the intended trial population, by giving more/less weight to patients when they are less/more likely (on the basis of their characteristics) to have the outcome measurement(s) observed, which here equates to the intercurrent events handled via the hypothetical strategy not occurring. 

The proposed estimator is consistent, provided that the model for not having a relevant intercurrent event (no missingness) is correctly specified, MAR allowing for time-varying covariates --observed outcomes as well as auxiliary variables-- holds and the positivity assumption is met. The MAR assumption requires that at each time in the trial, we have access to all prognostic factors (possibly time-varying) of outcome that are associated with having a relevant intercurrent event (e.g., of type 6 for Estimand 1). This is a weaker MAR assumption than the one considered for  the particular implementations of likelihood-based analysis and multiple imputation we considered earlier as we are allowing for time-varying covariates. The positivity assumption on the other hand states that the probability of not having a relevant intercurrent event (e.g., of type 6 for Estimand 1) given the observed history is always positive.

A disadvantage of this approach, which we discuss in more detail in the next section, is that other approaches can more easily deal with nonmonotone (i.e., intermittent) missing data.
Another important limitation of IPW estimators is their lower efficiency compared to likelihood-based and imputation approaches \citep{clayton1998analysis, carpenter2006comparison}. In the next section, we will show how the IPW estimator can be augmented by making direct use of the data from incomplete cases in order to improve upon its efficiency.

\subsubsection{Augmented Inverse Probability Weighting (AIPW)}\label{sec:aipw}

This class of estimators involves augmenting the simple inverse probability weighted complete case estimating equation \citep{robins1995analysis}. AIPW can be seen as a hybrid of IPW and imputation as it makes use of a model for the missingness mechanism and a model for the distribution of the complete (outcome) data. Estimators in this class therefore can yield improved efficiency compared to IPW estimators and have the appealing property of being doubly robust. The latter means that the estimator offers the analyst two chances, instead of only one, to obtain valid inference; an AIPW estimator is doubly robust if it is guaranteed to be consistent when either an imputation model (i.e., a model for the distribution of the complete (outcome) data) or a model for the probability of complete data is correctly specified by the user, but not necessarily both. This improved robustness is a crucial property not shared by standard IPW estimators or standard likelihood-based estimators. 

For simplicity, we first consider a trial where data are recorded at three time points: baseline ($t=0$), an intermediate time point ($t=1$) and the final time point (e.g., after 24 months; $t=2$). 
We can distinguish three cohorts of patients: a first cohort of patients for whom we have observed all data (i.e., all data are unaffected by the relevant intercurrent events; $\mathbf{C}^*_2=\mathbf{0}$), a second cohort of patients for whom we have observed the baseline measurements $X_0$ and $Y_0$ and the intermediate measurements $X_1$ and $Y_1$ ($\mathbf{C}^*_1=\mathbf{0}$ and $C^*_2=1$), and a last cohort of patients for whom only the baseline measurements $X_0$ and $Y_0$ are unaffected/observed ($C^*_1=1$ and $C^*_2=1$).
For trials with longitudinal follow-up and longitudinal monotone missing data, there are a number of approaches that can be taken \citep{bang2005doubly}.  
We review one specific AIPW imputation estimator for $E\left[Y_2(A=1, \mathbf{C}^*_{2}=\mathbf{0})\right]$ that has this doubly robust property and can be obtained using standard software via the following approach: 
\begin{enumerate}
\item At each time point $t$ ($t\in\{1,2\}$), we estimate the conditional probability  
$$P(C^*_t=0|A, \mathbf{\bar{C}_{t-1}}, \mathbf{\bar{X}_{t-1}}, \bar{Y}_{t-1})$$
by fitting a logistic regression among the patients with $\mathbf{C}^*_{t-1}=\mathbf{0}$.
\item Fit a generalized linear model with canonical link (e.g., linear regression for a continuous endpoint) for the outcome $Y_2$ among the treated ($A=1$) complete cases (cohort 1) given $\mathbf{\bar{X}_1}$ and $\bar{Y}_1$, using weights 
$$\prod_{t=1}^2\frac{1}{P(C^*_t=0|A, \mathbf{\bar{C}_{t-1}}, \mathbf{\bar{X}_{t-1}}, \bar{Y}_{t-1})}.$$
\item Use this model to impute $Y_2$ (using conditional mean imputation) for the treated patients in cohort 1 and 2 based on their observed measurements for $\bar{X}_1$ and $\bar{Y}_1$; Specifically, let $\hat{Y}_{i2}(\bar{X}_{i1}, \bar{Y}_{i1})$ denote the fitted value for patients $i$ for whom no missing data is observed up to at least time $1$ ($C^*_1=0$; patients in cohort 1 and 2). 
\item Fit a generalized linear model with canonical link (e.g., linear regression for a continuous endpoint) for the prediction $\hat{Y}_2(\bar{X}_1, \bar{Y}_1)$ among the treated ($A=1$) patients in the imputed dataset (cohort 1 and 2; $C^*_1=0$) given $X_0$ and $Y_0$ using weights
$$\frac{1}{P(C^*_1=0|A, X_{0}, Y_0)}.$$
\item Use this model to impute $Y_2$ (using conditional mean imputation) for \textbf{all} patients (all treated and untreated patients) based on their observed measurements for $X_0$ and $Y_0$. Specifically, let $\hat{Y}_{i2}(X_{i0}, Y_{i0})$ denote the fitted value for patient $i$.
\item Take the sample average of the fitted values $\hat{Y}_2(X_0, Y_0)$ over \textbf{all} patients.
\end{enumerate}
An estimator for $E\left[Y_2(A=0, \mathbf{C}^*_{2}=\mathbf{0})\right]$ can be obtained in a similar way but by reversing the role of treated and untreated patients.
The standard error can be estimated via the non-parametric bootstrap.

The algorithm for the neuroscience trial, where there are $9$ timepoints, is described in Section \ref{aipw_neuro} of the Supplementary Material.

In addition to monotone MAR missingness, this estimator relies on the following positivity assumption: at any time point $t$, any patient without missing data at time point $t-1$ has a positive probability of observing the outcome measurement at time $t$ given treatment assignment, baseline covariates (possibly including a pandemic time point indicator) and all post-baseline measurements up to time $t-1$. Under these assumptions, the resulting estimator is doubly robust, that is, it is consistent when the models for the probability of non-missingness are correctly specified at all time points, even when the outcome regression/imputation models are not, and vice versa \citep{bang2005doubly, hernan2020}. In addition, the proposed estimator is most precise (out of all inverse probability weighting estimators) if all models are correctly specified. In particular, it has the additional advantage of being more efficient than the IPW estimator because of the additional information in the outcome/imputation models. An algebraically equivalent sequential regression estimator is discussed in \citet{bang2005doubly}. This different representation leads to a computational algorithm that can be (more) easily implemented using standard off-the-shelf regression software. It moreover directly links with the targeted minimum loss-based estimation (TMLE) framework \citep[see e.g.,][]{van2012targeted}, which is the current frontier for embedding machine learning in the construction of doubly robust estimators. 
The development of coherent missing data models to account for missingness by IPW and AIPW has mainly been restricted for use in monotone MAR settings. To handle nonmonotone missingness, we refer the reader to \cite{vansteelandt2007estimation}. More recently, \cite{sun2018inverse} proposed a class of models for nonmonotone missing data mechanisms. Moreover, \cite{sun2018semiparametric} developed a semi-parametric estimation approach for MNAR data with the aid of instrumental variables.\\

\subsubsection*{Assumption `free' estimator}

\begin{figure}[ht]
	\centering
	\includegraphics[width=\textwidth]{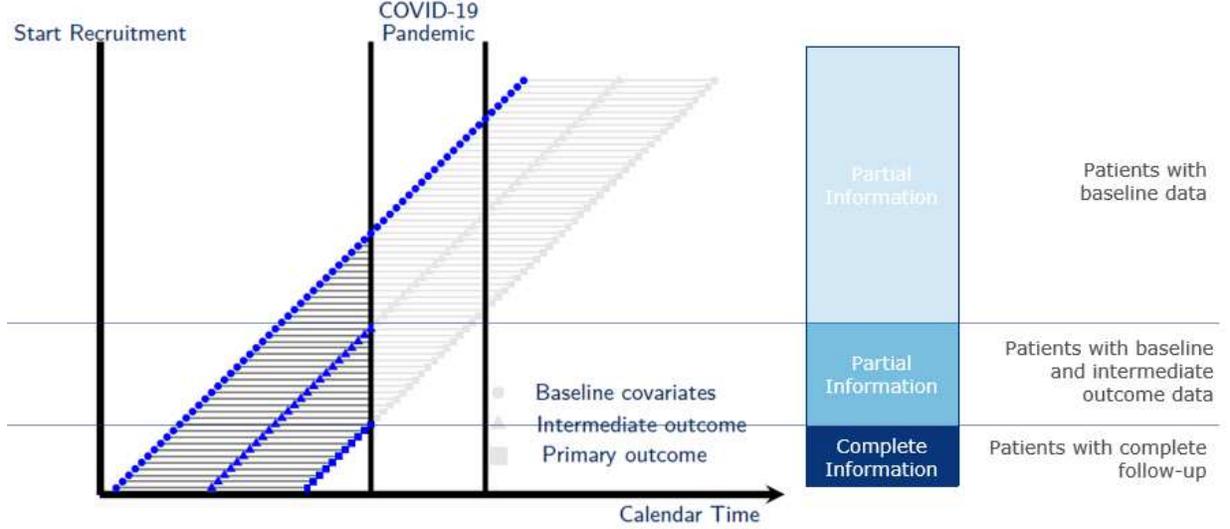}
	\caption{
		Visualization of the available data in a simplified setting with only three time points, assuming there is no population shift. The blue symbols show the measurements not impacted by COVID-19, while the gray symbols show the measurements (possibly) impacted by COVID-19.
	}
	\label{fig:covid_impacted_2}
\end{figure}

The AIPW estimator described earlier naturally leads to an ``assumption free'' estimator for the treatment effect in the absence of COVID-19, that is, in a world where the disease does not exist (Estimand 3). To see this, consider Figure \ref{fig:covid_impacted_2}, where the observed and unaffected data for the analysis is limited to the observed pre-pandemic data and (possibly) baseline data of patients recruited after the COVID-19 outbreak. Specifically, under the assumption that recruitment occurs randomly (i.e., independent censoring holds, in the sense that the probability of belonging to a certain cohort is fixed and does not depend on the (time-varying) covariates or outcomes) and that there is no population shift (i.e., patients recruited before, during and after the pandemic/COVID-19 are similar with respect to their baseline covariates),
we will also exploit the baseline measurements of the patients recruited post COVID-19 outbreak. Note that, for didactic reasons, this is a simplified setting with only one intermediate outcome. 
Although a reasonable amount of data might be thrown away and this estimator should therefore only be part of the sensitivity analysis, limiting the analysis to these data has the advantage that it overcomes possible misclassification of COVID-19-related intercurrent events. 

As before, we will consider the estimation of the treatment effect in a world without COVID-19 as a missing data problem. In the simplified case of one intermediate time point, we can distinguish three cohorts of patients (see Figure \ref{fig:covid_impacted_2}): a first cohort of patients for whom all data are available (i.e., not impacted by COVID-19), a second cohort of patients for whom only the baseline and intermediate measurements can be employed and a third cohort of patients for whom we can only employ the baseline measurements of the outcome and covariates. Here, if we assume no population shift, the third cohort of patients also includes patients randomized post COVID-19 outbreak. Note that, under random recruitment, the probability of belonging to a certain cohort is fixed and does not depend on the (time-varying) covariates or outcomes. The weighted regressions in the previous section can therefore be replaced with standard, unweighted regressions:
\begin{enumerate}
	\item fit a regression for the conditional mean of the outcome $Y_2$ given the baseline measurements $\mathbf{X_0}$ and $Y_0$ and the intermediate measurement $\mathbf{X_1}$ and $Y_1$ among the complete-cases (i.e., patients in cohort 1) in the treatment arm using a canonical generalized linear working model (with intercept),
	\item use this regression model to predict the outcome $Y_2$ for all patients in cohort 1 and 2 of the treatment arm, which we denote with $\hat{Y}_{i2}(\mathbf{\bar{X}_{i1}}, \bar{Y}_{i1})$,
	\item fit a regression of these predictions $\hat{Y}_{i2}(\mathbf{\bar{X}_{i1}}, \bar{Y}_{i1})$ on the baseline measurements $X_0$ and $Y_0$ among all patients in cohort 1 and 2 of the treatment arm using a canonical generalized linear working model (with intercept),
	\item use this regression model to predict the outcome $Y_2$ for all patients, which we denote by $\hat{Y}_{i2}(\mathbf{\bar{X}_{i1}}, Y_{i0})$; and taking the average of the predicted values $\hat{Y}_{i2}(\mathbf{\bar{X}_{i1}}, Y_{i0})$ over all patients (treated and untreated).
\end{enumerate}
In order to obtain an estimate of the treatment effect in a COVID-19 free world (Estimand 3), a similar approach can be followed for the average outcome under placebo (in a COVID-19 free world).

As we assume here the missingness mechanism (or the probability to belong to a certain cohort) is known, this estimator has the appealing feature of being asymptotically unbiased and consistent, even when the adopted prediction models are misspecified. A similar model-robust, covariate adjusted estimator has been used in \cite{benkeser2020improving} to improve precision in randomized trials for COVID-19 treatments.
However, misspecification will affect the estimator's variance, with the variance being minimized when the outcome/imputations models are correctly specified. 
Its variance can be estimated by the nonparametric bootstrap.

\subsection{Combining unbiased and possibly biased  estimators} \label{sec:est_combining} 
In this section, we present a statistical method to combine an unbiased (or asymptotically unbiased) estimator based on pre-pandemic data with another possibly biased estimator based on data collected during the pandemic. To combine these estimators into a single estimator with a smaller mean squared error or a smaller variance, we require existence of the first two moments of these estimators. These include many popular estimators, for example, maximum likelihood estimators, sample means, or  asymptotically normal estimates of regression coefficients. 

Methods for combining information from multiple sources to minimize mean squared error were proposed in \cite{Dmitriev2014} and \cite{tarima2020estimation}. Readers interested in Bayesian approaches for combining information from multiple data sources may refer to \cite{gravestock2017} and \cite{hobbs2012}; and for approaches on information borrowing see \cite{spiegelhalter2004bayesian}.

As it will be shown later, the optimal combined estimator works in the following circumstances. First, when the pandemic changes the effect of treatment, the impact of the data collected after the COVID-19 outbreak is (adaptively) suppressed by the combined estimator. Second, if the pandemic does not change the treatment effect, the data collected after the COVID-19 outbreak improves the accuracy of the estimator based on pre-pandemic data only.
Consider estimation of $\theta_{3}$ (denoted by $\theta$ in what follows) defined by Equation \eqref{estimand3a}, and assume for simplicity that $\theta=0$. When a trial is designed, the plan is to enroll $n$ patients and use an estimator $\hat\theta_n$ based on $n$ patients. Typically, this estimator is either unbiased $E(\hat\theta_n)=0$, or asymptotically unbiased by a central limit theorem. In the neuroscience example, the averaged CDR-SB score (the primary outcome) is assumed to be an unbiased estimator of $\theta$. When the pandemic strikes, only the first $m$ patients (i.e., the patients in cohort 1 in Section \ref{sec:aipw}) have their primary endpoint observed and the estimator $\hat\theta$ for $\theta$ with estimated variance $\widehat{var}(\hat\theta)$ can be calculated based on the pre-pandemic data. The primary endpoint collected for the remainder of the patients (i.e., patients in cohorts 2 and 3 in Section \ref{sec:aipw}) is summarized by $\tilde\theta$ with estimated variance $\widetilde{var}(\tilde\theta)$. As discussed in previous sections, the impact of the pandemic may change the data collection procedure, treatment adherence, and/or the treatment itself, which may or may not affect the distribution of the primary endpoint. 
It is reasonable to assume that the pandemic might lead to a biased estimator, $E(\tilde\theta)=\theta - \delta$. For example, in the neuroscience trial, as older and more fragile patients may not show up at some follow up visits, the averaged CDR-SB score over pandemic visits can be a biased estimator of $\theta$. It is possible, however, that $\delta = 0$ in which case the endpoint collected during the pandemic may improve estimation accuracy. This raises the question ``How to use the data collected during the pandemic without damaging the asymptotic properties of the estimator?''

Consider a class of linear combinations of the estimators $\hat\theta$ and $\tilde\theta$,
\begin{equation}\label{class_of_le}
\theta^{\lambda} = \hat\theta + \lambda\left(\hat\theta - \tilde\theta\right) = 
\hat\theta + \lambda\hat \delta,
\end{equation} 
where $\hat\delta=\hat\theta - \tilde\theta$.
The smallest mean squared error (MSE) in the class $\theta^{\lambda}$ is reached when $\lambda$ is equal to $\lambda_0:=-var(\hat\theta)E^{-1}(\hat\delta^2)$, where $E(\hat\delta^2) = var(\hat\theta) + var(\tilde\theta) + \delta^2$. Thus, the estimator with the smallest MSE is
\begin{equation}
\label{optest_MMSE_mult} 
\theta^{\lambda_0}(\delta) = \hat\theta - var(\hat\theta) E^{-1}(\hat\delta^2) \hat\delta
\end{equation}
with 
$$MSE[\theta^{\lambda_0}(\delta)] = var(\hat\theta) - 
[var(\hat\theta)]^2E^{-1}(\hat\delta^2).$$ 
Note that $\theta^{\lambda_0}(\delta)$ and its corresponding true MSE, $MSE[\theta^{\lambda_0}(\delta)]$, depend on $\delta$ as $E(\hat\delta^2)$ depends on $\delta$.
The unknown variance and expectation in $\theta^{\lambda_0}(\delta)$ are estimated leading to a new estimator $\theta^{\hat\lambda_0}(\delta)$, where $\hat\lambda_0 = -\widehat{var}(\hat\theta)\hat E^{-1}(\hat\delta^2)$.

In the special case of $\delta = 0$, $\theta^{\lambda}$ is unbiased for all choices of $\lambda$. Then, 
\begin{equation}
\label{optest_MVAR_mult} 
\theta^{\lambda_0}(0) = \hat\theta - var(\hat\theta)var^{-1}(\hat\delta)\hat\delta
\end{equation}
has the smallest variance among all $\theta^{\lambda}$ \citep[see][]{Tarima2006}, and 
\begin{eqnarray}
var[\theta^{\lambda_0}(0)] &=& var(\hat\theta) -
\frac{[var(\hat\theta)]^2}{ var(\hat\theta) + var(\tilde\theta)}.
\label{MSEoptest} 
\end{eqnarray}
The quadratic term at the right hand side of equation (\ref{MSEoptest}) shows the reduction in variance  when $\delta=0$. The estimators $\theta^{\hat\lambda_0}(0)$ and $\theta^{\lambda_0}(0)$ do not protect against the presence of bias. For example, if the pandemic impacts the treatment effect and $E(\tilde\theta)\ne \theta$, then $\theta^{\hat\lambda_0}(0)$ will be biased. Asymptotic relative efficiency of (\ref{optest_MVAR_mult}) is explored in \cite{Albertus2022}.

In practice, the estimators $\theta^{\hat\lambda_0}(\delta)$ and $\theta^{\lambda_0}(\delta)$ cannot be used because of their dependence on the unknown $\delta$. When $\hat\delta$  is substituted for $\delta$,  a new estimator $\theta^{\hat\lambda_0}(\hat\delta)$ is obtained.
If $\delta \ne 0$,  $\theta^{\hat\lambda_0}(\hat\delta)$ continues to be an asymptotically unbiased estimator of $\theta$ and is asymptotically equivalent to $\hat\theta$.
On the other hand, if $\delta = 0$, we should be more precise: $\sqrt{n}[\hat\theta^{0}(\hat\delta)-\theta]$ is no longer a normal random variable anymore; but it continues to converge to a stationary distribution and can be used for more powerful hypothesis testing as compared to the unbiased estimator. 
A Monte-Carlo example reported in Appendix Section \ref{mc_bi_unbi} illustrates the benefit of $\hat\theta^0(\hat\delta)$ when $\delta=0$.

In contrast to many model-based approaches presented in this article, the combined estimator minimizing MSE does not require any model assumptions except for existence of two moments and (asymptotic) unbiasedness of the originally planned estimator. Thus, it can be safely applied for estimating Estimand 3, the estimand associated with the original trial objective. The combined estimator is asymptotically equivalent to the estimator based on pre-pandemic data only if pandemic changed the treatment effect, and will have a smaller MSE, a shorter confidence interval, and will lead to more powerful testing if the pandemic did not affect the treatment effect. Since the effect of the pandemic is typically unpredictable, it is not impossible that despite the pandemic effects on recruitment rates, variablity (SDs), attrition rates and duration of follow-up, the treatment effect may continue to be unaffected. Thus, the use of the combined estimator is a safe option as it protects against the undesirable impact of model misspecification.

The estimation procedure described in this subsection can also be applied to estimate other estimands. The only requirement is the ability to estimate the summary measure of interest without bias. Then, other possibly biased estimators can be used to improve the MSE  of the original estimator.

\section{Recommendations for future trials}
\label{sec:future}
In this section, we outline different considerations for designing future trials in the context of the COVID-19 pandemic. 
Amongst the hypothetical estimands discussed in this paper, Estimand 2 may be the most relevant primary estimand for a trial that will be conducted when the pandemic is (almost) over. Below we further elaborate on the different attributes of the target estimand in the COVID-19 context. 

\subsection{Estimand attributes}

New trials designed in midst of the pandemic face a situation where new drugs are developed and treatment effects have to be estimated in a heterogeneous and unstable future. Participants can enter the trial with prior COVID-19 history or develop COVID-19 during the trial. We follow the estimand framework and thinking process from \citep{ich2019} to define the scientific question of interest and summarize key aspects to consider in designing new trials in the midst of the pandemic to target the relevant treatment effect to be estimated. Harmonization of design principles for new trials will be key for alignment across sponsors developing drugs and regulatory agencies evaluating drugs in similar disease indications. To help construct the target estimand, we describe these considerations in relation to the estimand attributes as follows:

\begin{itemize}
    \item When defining the \textbf{trial population}, the inclusion/exclusion criteria of a future clinical trial should be explicit regarding the COVID-19 status and other risk factors. This means that the eligibility criteria should be clear if there are any restrictions for patients with active COVID-19 infection, which might depend on the targeted disease. When deciding whether to make any such restriction, some of the patient factors to consider include geographic regions for trial conduct, racial diversity of patient population, underlying disease conditions, and duration of trial.
    \item The \textbf{primary variable} for new trials should reflect clinical practice, an adequate measure of clinical benefit and  acceptance by regulatory agencies. In certain situations, where the pandemic can impact the reliability of measurements of the primary endpoint, there may be a need for surrogate endpoints that can reliably measure the clinical benefit, such as estimands at landmark time points or  at earlier time points as opposed to later time points.    
    \item Given that patients may experience COVID-19 and require medication to treat it, the \textbf{definition of the treatment} of interest has to be explicit whether such additional medication becomes part of the treatment of interest i.e., `experimental treatment' against a `control' or the `experimental treatment plus COVID-19 drugs as needed' against `control plus COVID-19 drugs as needed'.
    Given COVID-19 medications may continue to be standard in a future pandemic-free world (e.g., when WHO declares the pandemic as endemic but still relatively many people may be infected and need such treatments), including COVID-19 drugs with experimental treatment and control as part of treatment of interest might be most relevant as it quantifies the treatment effect in a setting including the expected impact from medications used to treat COVID-19. 
    \item New trials are still likely to face additional challenges that stem from patient exposure to COVID-19, administrative lockdowns in specific regions and patient health impacts from COVID-19. Hence, defining relevant \textbf{intercurrent events} related to disease, pandemic-related lockdowns and concomitant medications used to treat COVID-19 and corresponding strategies to handle them will be key in addressing the scientific question of interest. For the estimand of interest, different strategies to deal with COVID-19-related intercurrent events versus other intercurrent events can be considered. For example, in the case of dose interruptions due to COVID-19 mobility restrictions, a hypothetical strategy could be considered, whereas withdrawal due to adverse events (whether COVID-19-related or not) can be dealt with by the treatment policy strategy or the composite strategy.
    
    \item The \textbf{summary measure} 
    depends on the expected (planned) duration of the trial and the reliability of the treatment effect estimate. 
\end{itemize}

\subsection{Design, Analysis Approach and Estimation}\label{subsec:designAnal}

There are further considerations to take into account when designing a trial during the pandemic. First, depending on the target population we may need to consider stratification, and enrichment based on prior COVID-19 status for an adequate representation of the patient population given the pandemic situation in the world. In particular, certain regions or countries may have different rates of COVID-19 infections and vaccination coverage, making patients more or less susceptible to suffering pandemic-related intercurrent events. Thus, at first, recruiting patients in areas with a lower risk can be more appealing in an effort to avoid the occurrence of pandemic-related intercurrent events. However, each national regulatory agency would like to have data that are representative for their own population. This may then lead to the need to recruit (more) patients from higher risk areas if these are relevant for the particular agency. 

Second, the time expected for the trial to be completed is crucial to anticipate whether it will occur before the pandemic is under control. If we expect that the trial will extend beyond this point, we will encounter a similar situation as the one depicted in Figure \ref{fig:covid_impacted}, except we will no longer have a pre-pandemic recruitment cohort. 

Unfortunately, the lines of demarcation between these cohorts are blurred. This is because besides requiring a clear definition of what a `post-pandemic' world entails, there is the potential for a `carry-over' pandemic effect. For example, even if patients recruited after a particular time are not at risk for COVID-19, those with a history of COVID-19 may still suffer effects from the disease. Excluding these patients from the trial population may in turn be problematic if the target population corresponds to a group with a substantial prevalence of COVID-19 history. The same issue arises when including patients who have received a vaccine.

If it is possible to rule out a carry-over effect, the next issue to consider is the proportion of patients that would need to be recruited after the pandemic is under control to make the results valid, as this would be the basis group to use for the imputation model. Another planning challenge is not knowing if and when the pandemic is going to be under control. All of these are important considerations and remain open questions for challenges for future trial design.

When the trial is planned, the true impact of the pandemic is typically unknown but it is possible that the impact is minimal. Then, estimation of estimands defined for the post-pandemic world can benefit from the data collected during the pandemic. Similarly, estimation of estimands defined for the pandemic affected world (`during pandemic') can benefit from the data collected during the post-pandemic phase. Researchers are typically reluctant to use  data collected in a `wrong' world, but are much more reluctant to exclude the data they have already collected. 
The approaches detailed in Section \ref{sec:estimation} can be applied in future trials to target for instance Estimand 2, where we envisage a setting where the intercurrent event disruptions were prevented from happening. As described in Section \ref{sec:missing}, we highlight the importance of carefully considering the missingness mechanism when choosing a statistical method for a particular context. The plausibility of the MAR assumption versus MNAR will determine the suitability of the method. Likewise, considerations regarding the need to account for time-varying covariates is warranted to aid the estimator choice. 
The estimator suggested in Section \ref{sec:est_combining} combines both data sources, continues to be asymptotically unbiased, and thus also is an attractive impartial solution to the researchers' dilemma. We refer the interested reader to Section \ref{mc_bi_unbi} for an illustration based on a Monte-Carlo trial.

Finally, the primary variable and other relevant information is often assessed at periodic visits during follow-up. Due to lockdowns and mobility restrictions, patients may miss one or more of their scheduled visits resulting in missing data. When planning future trials, ways to mitigate the impact of pandemic-related policies should be considered. To avoid missing data in this context, one consideration is to reduce the frequency of the visits or to have the assessments performed at home by a nurse. For certain outcomes, it may be possible to consider phone calls or telemedicine assessments, which, however, may not be available to all patients. Any modification of the above aspects need to take into account whether the treatment effect can be reliably estimated. For a detailed discussion on the use of remote assessments and other decentralized clinical trial elements we refer the reader to \citet{Alemayehu2021Decentralization}. Further considerations regarding the need of centrally adjudicated assessments can be considered. These considerations are relevant when planning data collection and data analysis. A possible drawback is the reliability and concordance of the measurements obtained by these alternative procedures. In addition, decentralized trials can introduce heterogeneity in data collection, which can result in changes in patient population and/or treatment management and which need to be accounted for when estimating the treatment effect of interest.\\

\section{Open Questions and Discussion}
\label{sec:recom}
In this article, we introduced different hypothetical estimand strategies for trials interrupted by COVID-19. Nevertheless, the estimands in Section \ref{sec:estimands} are not an exclusive list of estimands that may be of interest with respect to COVID-19 and the new COVID-19-related intercurrent events. For example, a variant on Estimand 2 is focusing on the treatment effect  in the patient population in the during- and post-pandemic world.
In addition, due to the continuing appearance of new COVID-19 variants, the treatment policy strategy becomes another important mechanism for handling some intercurrent events such as COVID-19-related treatment discontinuation. For handling such intercurrent events using a treatment policy strategy, some of the presented methods may not be applicable anymore.
Similarly, in case of a high frequency of intercurrent events the treatment policy might also be more appropriate than a hypothetical strategy for reliable estimation. 

Moreover, the occurrence of death due to COVID-19 will lead to additional challenges when formulating an estimand of interest. Death as an intercurrent event will require careful consideration and the appropriate strategy depends on the disease under study and the clinical endpoint \citep{meyer2020statistical,akacha2020challenges}.
For example, in disease areas with minimal mortality and where death is not a component of the primary endpoint, a hypothetical strategy for deaths related to COVID-19 infections
may be considered. However, the relevance and acceptability of a hypothetical estimand strategy seems to be less clear in more severe diseases where death due to other causes is part of the endpoint.
For example, how should we
account for deaths due to COVID-19 in a cardiovascular outcome trial where death is an outcome of
interest? Should this be
counted as an additional event for the outcome of interest or
addressed with an appropriate choice of strategy as an intercurrent
event? The impact of such intercurrent events, which may
occur at different rates in the different treatment arms, needs
to be carefully assessed on a case-by-case basis, and will in all likelihood lead to estimands not considered in this paper.


The estimands proposed in this article are defined to reflect the changes in the population across various time periods relative to the pandemic. Estimand 2, for instance, requires a definition of the time point at which COVID-19 is brought  under control. Although the demarcation separating the time  before the COVID-19 outbreak and the start of the pandemic is relatively clear, unfortunately, how to mark the start of the post-pandemic cohort is less apparent (see Section \ref{subsec:designAnal}). As different definitions of these cohorts are possible, we recommend defining them based on discussions with clinical experts for each trial -- depending on the disease area -- separately. Interest may, for instance, lie in a (post-pandemic) world where in every quarter at most 5\% of the patients get infected, or alternatively a (post-pandemic) world where at least 90\% of the patients are fully vaccinated. Note that for multi-site or international trials, cohort demarcation times  may vary at a site or country level and  may be informed by local COVID-19 infection rates. A more formal definition of the post-pandemic population might be considered once the pandemic has been re-classified as endemic by the WHO. The post-pandemic population might then be defined as the population (recruited) after the date set by the WHO to declare the end of the pandemic. The issues described for future trials regarding the potential for a `carry-over' effect also apply for ongoing clinical trials.

Moreover, not all study objectives can be addressed with sufficient quality. That is, in some cases we may not be able to estimate the targeted treatment effect reliably under plausible assumptions. For example, if an estimand targets a world without COVID-19, imputation models for predicting hypothetical values will be unreliable if the pre-COVID-19 sample sizes are too small. In addition to having large standard errors of regression coefficients, low quality imputation models may also miss important model predictors and include irrelevant predictors leading to problematic (multiple) imputations and possibly misleading conclusions. It is possible that the analytic model could become inestimable for some imputed datasets. The higher the percent of inestimable configurations of imputed datasets, the higher bias can be. Various sensitivity analyses can partially mitigate the impact of problematic imputations. Thus, the estimand choice must be balanced with practical considerations of what can be reliably and robustly estimated in present circumstances.

When the sample (of recruited patients) represents a different population than the original target population, results may not be (directly) generalizable to that original target population. As it happened with COVID-19, older patients were less likely to attend the doctors office, which shifted the demographics of recruited patients. For example, it might be that no patients $65$ years of age or older are recruited after the outbreak. Age is just one example of a pandemic's impact on  sampling; other examples include but are not limited to a patient's vaccination status and various comorbidities.
Different estimands can be considered when interest still lies in the target population. For example, if interest lies in a world without COVID-19, we suggest Estimand 3. Another option would be to estimate the effect in a post-pandemic world for the target population, by transporting data from the `post-pandemic population' to the `pre-pandemic population'. This can be accomplished using an outcome model and a propensity score model for the probability of belonging to the pre-COVID-19 population given baseline characteristics \citep{van2021efficient}. 
Because of the shift in (sample) population, it is also possible that interest lies in  an underrepresented subgroup after the pandemic, such as patients older than 65, rather than the entire original target population.  In this case, an analysis could be done to identify the risk for the patients older than 65 based on the available data prior to the COVID-19 outbreak. Such an analysis would be similar to an analysis for Estimand 3 for the population of patients older than 65.


In this article, we also have discussed different missing data methods, which provide an accessible route to addressing intercurrent events with the hypothetical strategy. However, they do necessitate strong assumptions about unobserved data in the hypothetical scenario of interest under MAR or MNAR. When post-intercurrent event  data are assumed to be similar to the data of observed individuals in the hypothetical scenario a MAR assumption may be most plausible, but critically MAR (or MNAR) is untestable. Therefore sensitivity analyses under alternative assumptions, also relevant for the estimand of interest, for any missing data should be conducted. Different missing data methods also entail additional assumptions as explored in Section \ref{sec:missing}. For example, an MI analysis requires the imputation model to be correctly specified and a proper method of MI be employed. In the context of COVID-19, decisions regarding what data is and is not relevant for the analysis (dependent on estimand strategies) are first required before any missing data methods can be applied. However, an advantage is that all observed data that are not impacted by the pandemic, including that which is obtained during the pandemic era, is used within the analysis if relevant for the estimand.

Finally, while we have proposed several estimands in this article, and yet others can be considered depending on the situation at hand, it is not clear to what extent estimates from these estimands will differ from each other in terms of their actual numerical values. For example, if the impact of the pandemic on our clinical trials is less than anticipated then the different estimands could lead to similar treatment effect estimates. To investigate how different the various estimands are from each other, we should  conduct data analyses and run simulations when we can take advantage of data from disrupted trials in different disease areas. Specifically, we should explore what is driving the possible differences in magnitude and direction. Disease area may be a key factor influencing the impact of the pandemic -- and it might be different in various countries and at different points in time. The hope is that we will be able to learn from the COVID-19 experience in order to better handle future trial interruptions.

\section*{Acknowledgements}
The authors thank the National Institute of Statistical Sciences for facilitating this work on Estimands and Missing Data, which is part of the Ingram Olkin Forum Series on Unplanned Clinical Trial Disruptions. The authors would also like to recognize the organizers of this forum series (those who are not an author on this paper): Chris Jennison and Adam Lane as well as the speakers at the motivating workshop (those who are not an author on this paper): Mouna Akacha and David Murray.

K.V.L. is supported by Fulbright Belgium, Belgian American Educational Foundation and VLAIO (Flemish Innovation and Entrepreneurship) under the Baekeland grant agreement HBC.2017.0219. 
S.T. is partially supported by the Department of Health and Human Services of the National Institutes of Health under award number R40MC41748.
J.B. and C.O.P. are supported by UK Medical Research Council under grant agreement MR/T023953/1.
H.M. is supported by VLAIO under Baekeland grant agreement HBC.2019.2155.
S.C. is supported by an NIHR advanced research fellowship (NIHR 300593). The views expressed are those of the authors and not necessarily those of Fulbright Belgium, Belgian American Educational Foundation, VLAIO, the National Institutes of Health, the UK Medical Research Council, the  NIHR or the Department of Health and Social Care.

We are grateful for feedback of Marcel Wolbers and Kaspar Rufibach on earlier versions of this manuscript.

\section*{SUPPLEMENTARY MATERIAL}

\subsection{AIPW estimator for the neuroscience trial}\label{aipw_neuro}

For the neuroscience trial (see Section \ref{sec:exam}), the algorithm to obtain a double robust estimator is as follows:
\begin{enumerate}
\item At each time point $t$ ($t\in\{1,\dots,8\}$), we estimate the conditional probability  
$$P(C^*_t=0|A, \mathbf{\bar{C}_{t-1}}, \mathbf{\bar{X}_{t-1}}, \bar{Y}_{t-1})$$
by fitting a logistic regression among the patients with $\mathbf{C}^*_{t-1}=\bar{0}$.
\item Fit a generalized linear model with canonical link (e.g., linear regression for continuous endpoint) for the outcome $Y_8$ among the treated ($A=1$) complete cases given $\mathbf{\bar{X}_7}$ and $\bar{Y}_7$, using weights 
$$\prod_{t=1}^8\frac{1}{P(C^*_t=0|A, \mathbf{\bar{C}_{t-1}}, \mathbf{\bar{X}_{t-1}}, \bar{Y}_{t-1})}.$$
Let $\hat{Y}_{i,8}(\mathbf{\bar{X}_{i,7}}, \bar{Y}_{i,7})$ denote the fitted value for patient $i$ (in the treatment arm) for whom no missing data is observed up to at least time $7$ ($\mathbf{C}^*_7=0$).
\item Recursively, for $t^*=7, \dots, 2, 1$: fit a generalized linear model with canonical link for $\hat{Y}_{i,t^*}(\mathbf{\bar{X}_{i,t^*-1}}, \bar{Y}_{i,t^*-1})$ among the patients with no missing data up to at least time $t^*$ given $\mathbf{\bar{X}_{t^*-1}}$ and $\bar{Y}_{t^*-1}$ using
$$\prod_{t=1}^{t^*}\frac{1}{P(C^*_t=0|A, \mathbf{\bar{C}_{t-1}}, \mathbf{\bar{X}_{t-1}}, \bar{Y}_{t-1})}.$$
Let $\hat{Y}_{i,t^*}(\mathbf{\bar{X}_{i, t^*-1}}, \bar{Y}_{i, t^*-1})$ denote the fitted value for patient $i$ (in the treatment arm) for whom no missing data is observed up to at least time $t^*-1$ ($\bar{E}^*_{t^*-1}=0$).
\item Take the sample average of the fitted values $\hat{Y}_8(X_0, Y_0)$ over \textbf{all} patients (treated and untreated).
\end{enumerate}

\subsection{A Monte-Carlo Study on Combining Unbiased and Possibly Biased Estimators.} \label{mc_bi_unbi} To illustrate the relative benefits of $\hat\theta^{0}(0)$ and $\hat\theta^{0}(\hat\delta)$ versus $\hat\theta$, consider an illustrative Monte-Carlo study with the statistical model generated two samples: (1) the pre-pandemic sample, which is a sample with $100$ paired standard normal random variables ($X_1$ representing the primary endpoint and $Y_1$ representing a surrogate endpoint) with correlation $cor(X_1,Y_1) = 0.9$ and (2) the pandemic sample with $1000$ standard normal random variables ($X_2$ - surrogate endpoint assessed during the pandemic). The objective is to estimate the mean of $Y$ ($EY=\theta$), which is equal to zero in this example.

\begin{figure}[h!]
     \centering
     \begin{subfigure}[b]{0.45\textwidth}
        \centering
        \includegraphics[scale=0.4]{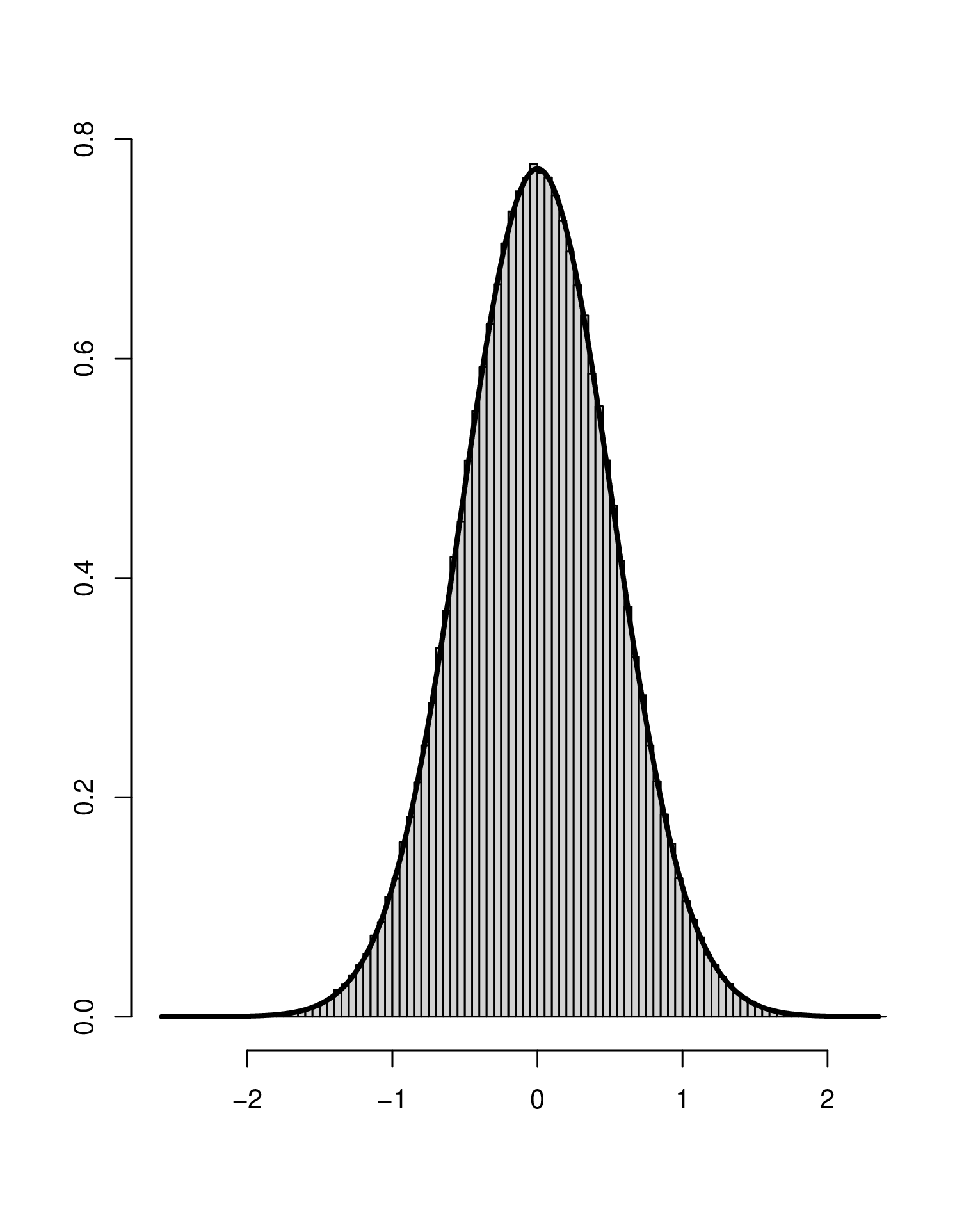}
        \caption{$\sqrt{100}\hat\theta^0(0)$}
        \label{fig2.1}
     \end{subfigure}
     \begin{subfigure}[b]{0.45\textwidth}
        \centering
        \includegraphics[scale=0.4]{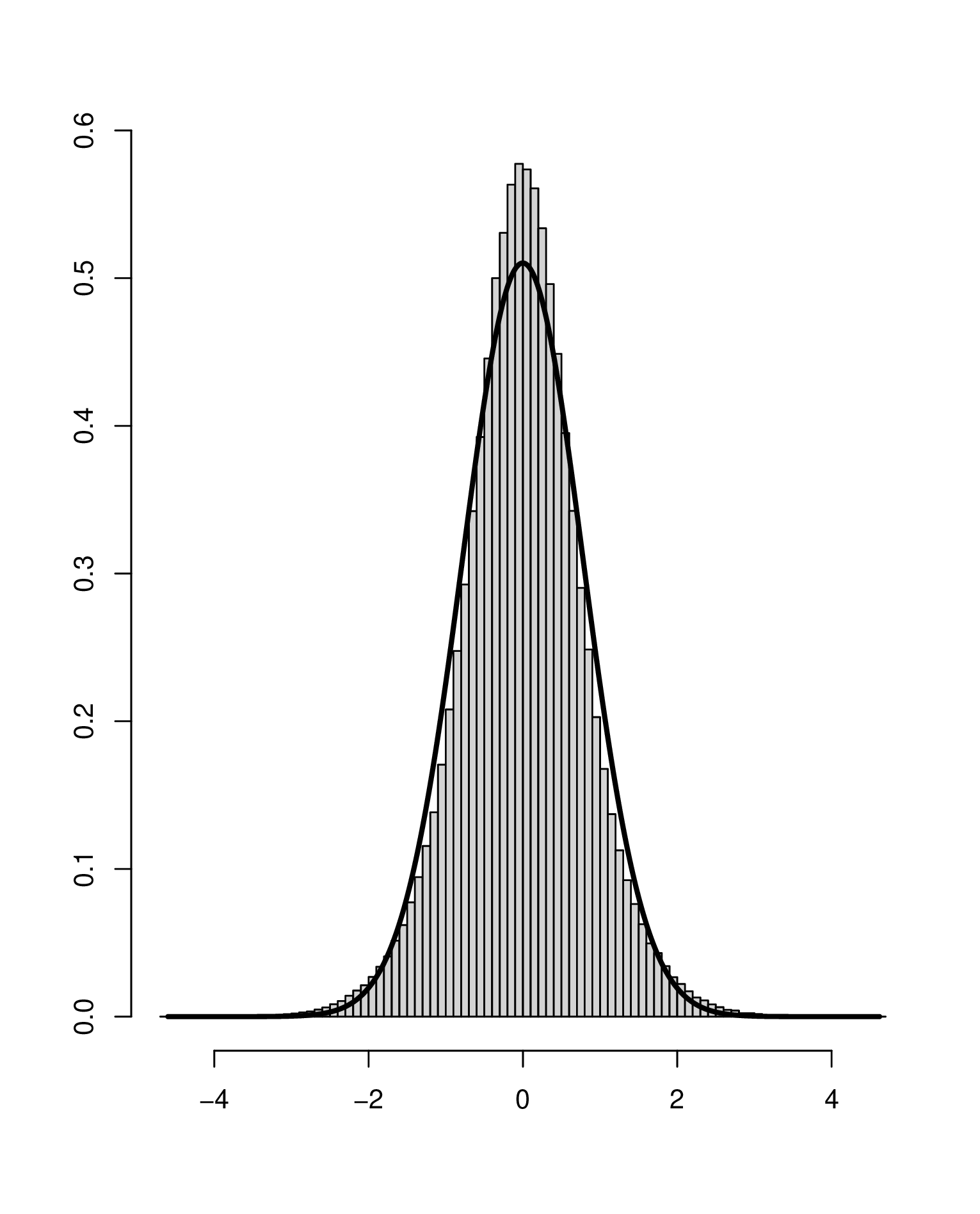}
      \caption{$\sqrt{100}\hat\theta^0(\hat\delta)$}
      \label{fig2.2}
    \end{subfigure}
    \caption{Histogram and a normal approximation of the distribution; $500,000$ Monte-Carlo   simulations.}
\end{figure}



The asymptotic distribution of $\hat\theta$ (mean of $Y_1$) is approximately normal, so that $\sqrt{100}\cdot\hat\theta\sim N(0,1)$ leads to the width of $95\%$ for $\sqrt{100}\cdot\hat\theta$ equal to $3.92 (=2\cdot1.96)$. The asymptotic distribution of $\sqrt{100}\cdot\hat\theta^0(0)$ is also approximately normal with zero mean and variance $=0.266358$, see Figure \ref{fig2.1}. The distance between $2.5\%$ and $97.5\%$ level quantiles of the distribution of $\sqrt{100}\cdot\hat\theta^0(0)$ is equal to $2.03227$. Wald's confidence interval (``mean estimate'' $\pm 1.96$ ``standard deviation of the estimate'') has an almost identical length ($=2.023107$).

The asymptotic distribution of $\sqrt{100}\cdot\hat\theta^0(\hat\delta)$ is not normal anymore and is shown in Figure \ref{fig2.2}. The normal approximation allows us to visually evaluate the departure from normality. The absence of asymptotic normality, however, is not really a problem. Since the asymptotic distribution is known it can be used for estimation, hypothesis testing, and for calculating confidence intervals. For example, the distance between the $2.5\%$ and $97.5\%$ level quantiles of the distribution of $\sqrt{100}\cdot\hat\theta^0(\hat\delta)$ is equal to $3.20191$. Wald's confidence interval has a shorter length ($=3.064861$) associated with a less than $95\%$ coverage.

This Monte-Carlo study demonstrates that if a data analyst is confident that pandemic data on a surrogate endpoint is unbiased, then it should be incorporated using minimum variance estimation ($\hat\theta^0(0)$). If, however, it can be biased, $\hat\theta^0(\hat\delta)$ is a more appropriate method. 

\bibliographystyle{chicago}

\bibliography{Bibliography-MM-MC}
\end{document}